\documentclass[review]{elsarticle}

\usepackage{graphicx}
\usepackage{subfig}
\usepackage{float}
\usepackage{epstopdf, epsfig}
\usepackage{csquotes}
\usepackage{natbib}
\usepackage{color}
\usepackage{amssymb,amsmath,mathtools}
\usepackage{bm} 
\definecolor{Pink}{rgb}{0.8,0.2,0.5451}

\usepackage{lineno,hyperref}
\modulolinenumbers[5]

\journal{Computers \& Fluids}

\begin{document}

\begin{frontmatter}

\title{Simulation of wake bimodality behind squareback bluff-bodies using LES}

\author[1]{F. Hesse\corref{CorAuthor}}
\ead{faron.hesse10@imperial.ac.uk}
\author[1]{A. S. Morgans}

\address[1]{Department of Mechanical Engineering, Imperial College London, London SW7 2AZ, UK}

\cortext[CorAuthor]{Corresponding author}

\begin{abstract}
A large eddy simulation (LES) study of the flow around a 1/4 scale squareback Ahmed body at $Re_H=33,333$ is presented. The study consists of both wall-resolved (WRLES) and wall-modelled (WMLES) simulations, and investigates the bimodal switching of the wake between different horizontal positions. Within a non-dimensional time-window of 1050 convective flow units, both WRLES and WMLES simulations, for which only the near-wall region of the turbulent boundary layer is treated in a Reynolds-averaged sense, are able to capture horizontal (spanwise) shifts in the wake's cross-stream orientation. Equilibrium wall-models in the form of Spalding's law and the log-law of the wall are successfully used. Once these wall-models are, however, applied to a very coarse near-wall WMLES mesh, in which a portion of the turbulent boundary layer's outer region dynamics is treated in a Reynolds-averaged manner as well, large-scale horizontal shifts in the wake's orientation are no longer detected. This suggests larger-scale flow structures found within the turbulent boundary layer's outer domain are responsible for generating the critical amount of flow intermittency needed to trigger a bimodal switching event. By looking at mean flow structures, instantaneous flow features and their associated turbulent kinetic energy (TKE) production, it becomes clear that the front separation bubbles just aft of the Ahmed body nose generate high levels of TKE through the shedding of large hairpin vortices. Only in the reference WRLES and (relatively) fine near-wall mesh WMLES simulations are these features present, exemplifying their importance in triggering a bimodal event. This motivates studies on the suppression of wake bimodality by acting upon the front separation bubbles. 
\end{abstract}

\begin{keyword}
Large eddy simulation (LES), wall-resolved LES, wall-modelled LES, wake bimodality, hairpin vortices, turbulent kinetic energy
\end{keyword}

\end{frontmatter}


\section{Introduction} \label{Intro}
Many road vehicles, such as lorries or light good vehicles, are classified as bluff-bodies with their under-body in close proximity to the ground. At their (rear) trailing edge, the boundary layer separates, thereby creating a substantial re-circulation bubble aft of the vehicle. This large re-circulation zone is characterised by low pressure and is called a wake. It contributes most to the vehicle's pressure drag, which is the dominating term of the vehicle's total aerodynamic drag \cite{RefWorks:210}. To effectively lower the aerodynamic drag exerted on the vehicle and, in turn, lower combustion powered vehicle emission or increase electric car range, the wake's dynamics must be well understood.

The time-averaged behavior and structure of the wake has been studied extensively. Possibly the earliest experimental work that gathered quantitative information on the time-averaged wake was done by Ahmed \cite{RefWorks:212}, who studied the separation bubble and pair of counter-rotating longitudinal vortices that characterize the wake behind estate, fastback and notchback rear-end geometries. Shortly afterward, a wake study was performed on the Ahmed body with varying base slants \cite{RefWorks:210}. Meanwhile, some of the first numerical work on the time-averaged wake was carried out by Han \cite{RefWorks:253}, who simulated the flow past the Ahmed body using the Reynolds-averaged Navier-Stokes (RANS) equations and the $k-\epsilon$ turbulence model. Further numerical work on the steady-state wake's topology for square-back geometries was performed by Krajnovi{\'c} and Davidson \cite{RefWorks:214}, who looked at the time-averaged flow results from large eddy simulations (LES) of a simplified bus shape. Their study clearly illustrates that the near wake's time-averaged toroidal structure interacts with the encompassing shear layers to generate the aforementioned counter-rotating vortices further downstream.

Recently, investigations have started to focus on the instantaneous dynamics of the rear-end wake. This is because a vehicle's aerodynamic properties, such as lift, drag, wind noise and general vehicle stability, are inherently related to the external flow field's transient motions \cite{RefWorks:214}. Grandemange et al. \cite{RefWorks:215} were the first to observe experimentally that the laminar three-dimensional wake aft of the Ahmed body bifurcates to a permanent state of broken reflectional symmetry once a critical Reynolds number based on body height, $Re_H$, of 340 is reached. Unsteadiness within the laminar wake initiates slight asymmetries that subsequently force the wake into an off-axis position. In the laminar regime, the asymmetric wake is locked into its current off-center location \cite{RefWorks:215}. These experimental results are supported by the work of Evstafyeva et al. \cite{RefWorks:207}, who were the first to resolve the wake bifurcation scenarios numerically for the Ahmed body.

It is well-established that the wake's broken symmetry persists in the turbulent $Re_H$ regime. This was illustrated experimentally for an axisymmetric bluff body by Rigas et al. \cite{RefWorks:299}. They found that the three-dimensional turbulent wake's large-scale coherent structures retain in a statistical sense the structure observed for the laminar symmetry-breaking instabilities. Unlike the laminar wake though, the turbulent wake exhibits an infinite number of symmetry-breaking states; it is multi-stable. For rectilinear three-dimensional bluff bodies, like the Ahmed body, this translates into a phenomenon known as bimodality; the asymmetric wake shifts randomly between two opposing off-axis locations. Bimodality generates an unsteady side force, which contributes to about 3-7\% of bluff body drag \cite{RefWorks:254}. Extensive studies on wake bimodality and its suppression have, therefore, been conducted. Grandemange et al. \cite{RefWorks:202} probably carried out some of the first experimental work on the Ahmed body's bimodal wake. In addition to identifying that bimodality appears to be independent of the turbulent $Re_H$ and that bimodality manifests itself only past a critical ground clearance value, Grandemange et al. \cite{RefWorks:202} conclude that the wake presents instantaneous dynamics over two characteristic time-scales: at short time-scales of $T_s \sim 5 H / |\bm{U}_{\infty}|$, where $|\bm{U}_{\infty}|$ is the free-stream velocity's magnitude, the wake exhibits coherent oscillations of mild strength in the lateral and vertical directions to give the vortex shedding mode, while over long random time-scales up to $T_l \sim 10^3 H / |\bm{U}_{\infty}|$ the wake presents large-scale horizontal shifts in the orientation of its toroidal structure to give the bimodal mode. Also, using two-dimensional PIV planes, theories on the unsteady wake's topology have been proposed: Evrard et al. \cite{RefWorks:217} suggest the instantaneous wake is made up of one horseshoe vortex system, while both Perry et al. \cite{RefWorks:216} and Pavia et al. \cite{RefWorks:329} conjecture that the time-dependent wake evolves via a multiple horseshoe vortex system. To re-centre the wake and, thus, raise vehicle base pressure, both Evrard et al. \cite{RefWorks:217} and Lucas et al. \cite{RefWorks:218} appended deep base cavities to the rear of the Ahmed body. Furthermore, feedback control using slot jets \cite{RefWorks:255} and oscillating flaps \cite{RefWorks:256}, respectively, successfully centred the wake and raised the Ahmed body's base pressure. It has also been found that a steady symmetric force, applied to the Ahmed body's wake using a vertical control cylinder, can re-centre and stabilize the wake, leading to an increased re-circulation bubble length and a concomitant base pressure recovery \cite{RefWorks:254}.

\begin{figure} 
\centering
\graphicspath{ {ComputationalSetUp/} }
\includegraphics[scale=0.45]{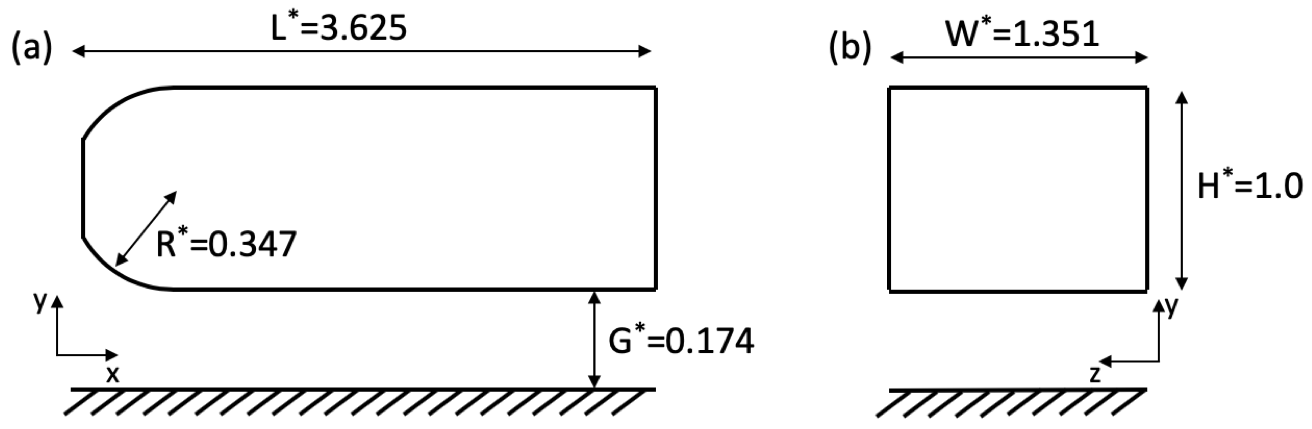}
\caption{Ahmed body schematic - (a) side view and (b) rear view}
\label{fig:AhmedBody}
\end{figure}

Most currently published work on understanding the turbulent asymmetric wake and its bimodal dynamics is experimental. It seems that only Pasquetti and Peres \cite{RefWorks:257}, Lucas et al. \cite{RefWorks:218}, Rao et al. \cite{RefWorks:300} and Rao et al. \cite{RefWorks:297} have been able to naturally resolve the turbulent wake's asymmetry numerically. More importantly, Dalla Longa et al. \cite{RefWorks:306} present the only computational study that features bimodality and this was accomplished using fully block-structured baseline mesh sizes of 56.6 million cells suggesting that highly resolved simulations are needed to capture the dynamics responsible for forcing a bimodal switch. Such large meshes preclude the execution of numerical based drag reduction studies centred on the bimodal wake. The purpose of this study is, therefore, twofold: (a) to create a computationally feasible numerical set-up that features bimodality for future simulation based drag reduction work, and (b) to extend the numerical work of Dalla Longa et al. \cite{RefWorks:306} by further identifying and analysing possible flow mechanisms responsible for forcing a bimodal switch. These two objectives are accomplished by simulating a $Re_H$=33,333 flow past the 1/4 scale squareback Ahmed body (figure \ref{fig:AhmedBody}) from Ahmed et al. \cite{RefWorks:210} with meshes of varying near-body resolution coupled to the use of wall-resolved LES (WRLES) or wall-modelled LES (WMLES). Comparative studies between WRLES and WMLES are not uncommon. Aljure et al. \cite{RefWorks:287}, for instance, assessed the benefit of using WMLES over WRLES for simulating the flow field around the DrivAer-fastback car model, concluding that WMLES reduces CPU time by 70\% while returning high-fidelity unsteady data.

The remainder of this paper is organized as follows. Section \ref{ComputatonalSetUp} describes the computational set-up, which includes a mesh refinement study and wall-model analysis. Meanwhile, section \ref{Results} illustrates the results and explains their significance. Finally, section \ref{Conclusion} concludes the paper and offers some perspective on future work.

\section{Computational set-up} \label{ComputatonalSetUp}

\subsection{Geometry and flow domain}
The 1/4 scale squareback Ahmed body in figure \ref{fig:AhmedBody} has the following dimensions: length of $L$=0.261m, height of $H$=0.072m, width of $W$=0.09725m, nose radius of $R$=0.025m and ground clearance of $G$=0.0125m. These dimensions match those used by Grandemange et al. \cite{RefWorks:202} in their seminal work on wake bimodality aft of the squareback Ahmed body. When non-dimensionalized by body height, $H$, the size specifications given in figure \ref{fig:AhmedBody} are obtained; non-dimensional size specifications are marked by an asterisk, $^*$.

For flows around bluff-bodies, ERCOFTAC recommends that the computational domain, $\Omega$, be approximately of size $\Omega = (L_{inlet},L_x,L_y,L_z) = (2L,8L,2L,2L)$, where $L$ is the bluff-body's length \cite{ERCOFTAC2}. This is the domain size adopted in the study.

\subsection{Numerical solver}
Simulations are performed using the dimensional open-source computational fluid dynamics (CFD) toolbox OpenFOAM that employs the finite volume method to solve the Navier-Stokes equations. The simulations are carried out in a two-part process. First, to accelerate the simulation process and, thereby, reduce the computational resources used, all simulations are initialized to a steady-state using the simpleFOAM solver, a time independent solver for incompressible turbulent flow that uses the semi-implicit method for pressure linked equations (SIMPLE) algorithm to solve for the coupled pressure and velocity fields. For this portion of the simulation procedure, the schemes chosen to solve the spatial derivatives are first-order accurate, as low numerical cost and robustness, instead of accuracy, are first targeted. As such, the RANS turbulence model selected for the initialization simulations is the Spalart-Allmaras model \cite{RefWorks:331}. Once the simulations reached a steady-state, the solution is propagated forward in time using the pimpleFOAM solver, a large time-step transient solver for incompressible turbulent flow that uses the pressure implicit with splitting operators (PISO) – SIMPLE algorithm for evaluating the coupled pressure and velocity fields. The time and spatial derivative solution schemes adopted for part two of the simulation process are second-order accurate, as accuracy is now targeted. To maintain solver stability by keeping the solution total variation diminishing (TVD) and, thereby, preventing the creation of overshoots and undershoots in the solution domain, the Courant-Friedrichs-Lewy (CFL) number is kept at a magnitude of \textit{O}(1) and limiters are applied to the solver schemes \cite{RefWorks:137}. The LES turbulence model applied is the wall-adapting local eddy-viscosity (WALE) model developed by Nicoud and Ducros \cite{RefWorks:229}. WALE offers advantages over the classical Smagorinsky model used in LES, such as the reproduction of the eddy-viscosity's cubic wall-asymptotic behavior, and accounting for both the strain and rotation rate of the finest resolved turbulent eddies, thereby better modelling the energy transfer from the resolved scales to the sub-grid scales \cite{RefWorks:229}.

\begin{figure}
\centering
\graphicspath{ {ComputationalSetUp/} }
\includegraphics[scale=0.43]{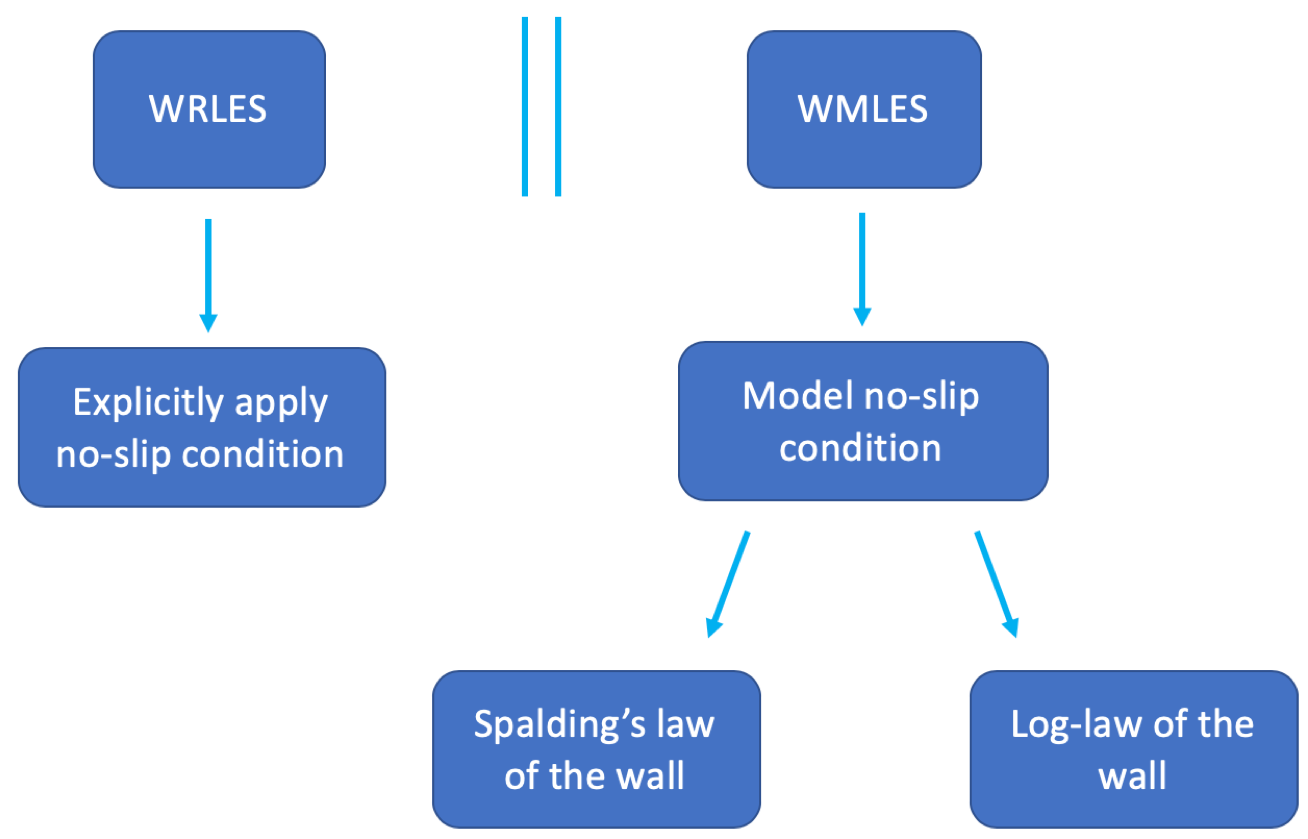}
\caption{Flow chart of selected wall boundary conditions}
\label{fig:WallBC}
\end{figure}

\subsection{Boundary conditions}
To simulate a $Re_H$=33,333 flow past the 1/4 scale squareback Ahmed body, where $H$=0.072m and fluid viscosity is $\nu$=1.511$\times$10$^{-5}$m$^{2}$s$^{-1}$ for air at 20$^{\circ}$C, a uniform flow velocity condition of $\bm{U}=(6.996,0,0)$ms$^{-1}$ without artificial forcing (i.e. turbulence develops naturally) is prescribed at the domain inlet. This corresponds to a free-stream velocity magnitude, $|\bm{U}_{\infty}|$, of 6.996ms$^{-1}$. Far away from the Ahmed body at the computational domain's sides and top upper boundary the free-slip condition is enforced. Meanwhile, a pressure advection condition is applied at the domain outlet. At the ground and Ahmed body walls, three types of boundary conditions are applied. The selected condition depends on whether WRLES or WMLES is performed, as illustrated in the flow chart of figure \ref{fig:WallBC}. For WRLES, the no-slip condition is explicitly enforced and the entire boundary layer is resolved. The required high near-wall mesh resolution leads to large mesh sizes (see section \ref{MeshRefineStudy}). If WMLES is carried out, the turbulent boundary layer's near-wall dynamics and, at most, a portion of its outer region physics are modelled by treating the flow in a Reynolds-averaged sense. Only (some of) the outer turbulent boundary layer's energetic eddies are captured by the mesh. With WMLES, coarser near-wall meshes are, therefore, permitted (see section \ref{MeshRefineStudy}). The wall-adjacent cell's flow information is fed into a wall-model that subsequently computes the wall shear stress, $\tau_w$, for that cell's wall boundary. A wall-model turbulent eddy viscosity, $\nu_{t,wm}$, is then calculated, which, coupled to the impermeability condition, leads to a modelled no-slip boundary condition. In this study, two RANS based wall-stress equilibrium models are adopted: Spalding's law of the wall (equation \ref{SpaldLaw}) and the log-law of the wall (equation \ref{LogLaw}). The following definitions are adopted for formulating both wall-models: $u_1^+=u_1/u_{\tau}$, $y_1^+=y_1u_{\tau}/\nu$, $u_{\tau}=\sqrt{\tau_{w}/\rho}$, $\kappa=0.42$ and $B=5.4$. Sub-script \enquote{1} denotes the wall-adjacent cell  \citep{RefWorks:260,RefWorks:231,RefWorks:289,RefWorks:326,RefWorks:323}.
\begin{equation}
    y_1^+ = u_1^+ + \text{e}^{-\kappa B} \Big[ \text{e}^{\kappa u_1^+} - 1 - \kappa u_1^+ - \frac{1}{2} ( \kappa u_1^+ )^2 - \frac{1}{6} ( \kappa u_1^+ )^3 \Big] \label{SpaldLaw}
\end{equation}
\begin{equation}
    u_1^+ = \frac{1}{\kappa} \text{ln}(y_1^+) + B \label{LogLaw}
\end{equation}
Spalding's law of the wall is valid for both the laminar viscous sub-layer and the log-region of the turbulent boundary layer. Meanwhile best-practice guidelines dictate that the log-law of the wall be, ideally, applied if the wall-adjacent cell's centre is positioned in the log-region of the turbulent boundary layer. This corresponds to $y_1^+$ values in the range of $30 \leq y_1^+ \leq 1000$ or higher.

\subsection{Definition of key variables}
\begin{figure}
\centering
\graphicspath{ {ComputationalSetUp/} }
\includegraphics[scale=0.45]{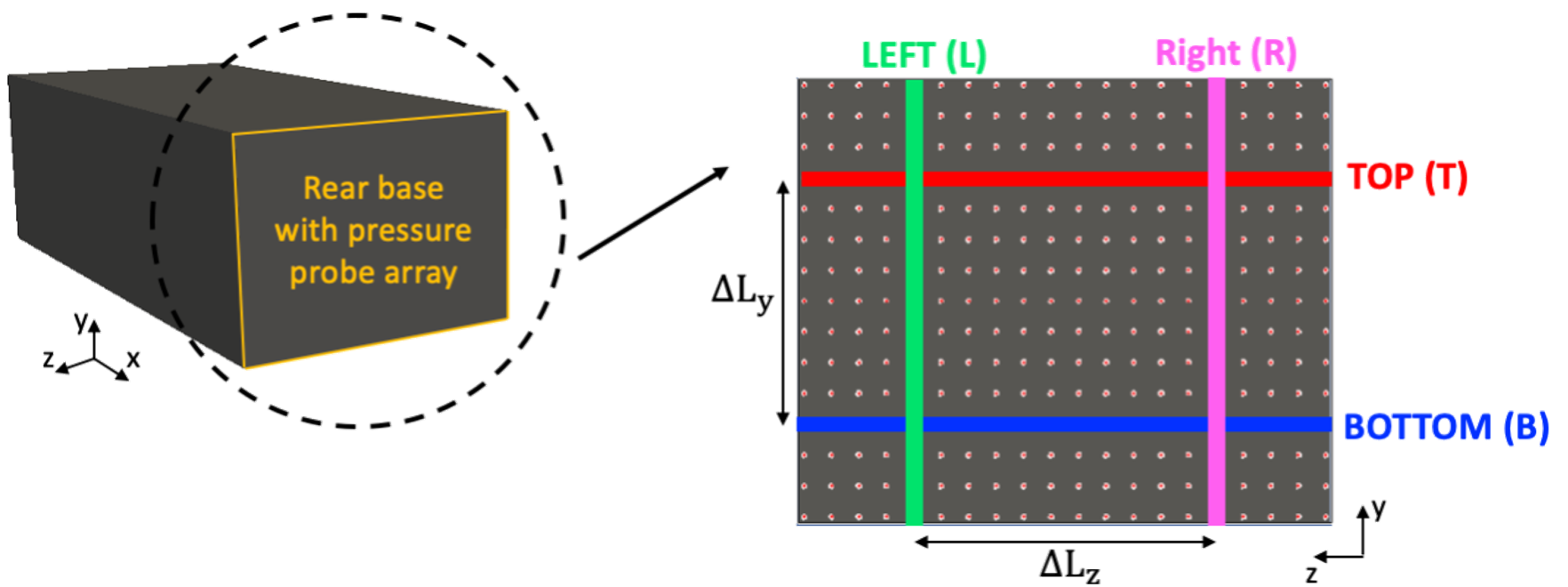}
\caption{Pressure array on the Ahmed body's rear base}
\label{fig:BasePressureArray}
\end{figure}

The following sub-section serves to define key variables that, in addition to pressure, $P$, and flow velocity, $\bm{U}$, are used throughout this investigation. Non-dimensional coefficients include the drag coefficient, $C_D$, the lift coefficient, $C_L$, and the pressure coefficient, $C_P$, from which the base pressure coefficient, $C_{PB}$, is computed for the area $A=W H$ using $n=300$ pressure probes (figure \ref{fig:BasePressureArray}) (equations \ref{Cdrag}-\ref{Cpressure}). These are defined below, where $F_D$ is the drag force, $F_L$ is the lift force, $P_{\infty}$ is the free-stream pressure, $\rho_{\infty}$ is the free-stream density, $A_{yz}$ is the area of the Ahmed body projected onto the $yz$-plane and $A_{xz}$ is the area of the Ahmed body projected onto the $xz$-plane.
\begin{equation}
    C_D=\frac{F_D}{\frac{1}{2} \rho_{\infty} |\bm{U}_{\infty}|^2 A_{yz}} 
    \label{Cdrag}
\end{equation}
\begin{equation}
    C_L=\frac{F_L}{\frac{1}{2} \rho_{\infty} |\bm{U}_{\infty}|^2 A_{xz}} 
    \label{Clift}
\end{equation}
\begin{equation}
    C_P=\frac{P-P_{\infty}}{\frac{1}{2} \rho_{\infty} |\bm{U}_{\infty}|^2 } \quad \rightarrow \quad C_{PB}=\frac{\sum_{i=1}^{n=300} C_{P_i} \mathrm{d}A}{W H} 
    \label{Cpressure}
\end{equation}
Quantities of crucial importance to this study are the Ahmed body rear base pressure gradient in the vertical direction, $\partial C_P / \partial y$, and the horizontal direction, $\partial C_P / \partial z$. These are defined with reference to figure \ref{fig:BasePressureArray}, which depicts the entire pressure probe array on the Ahmed body base and, more importantly, four distinct pressure probe arrays: two horizontally oriented pressure probe arrays that are marked as TOP (T) and BOTTOM (B), and two vertically oriented pressure probe arrays that are marked as LEFT (L) and RIGHT (R). $\partial C_P / \partial y$ is based on the delta between the average of the TOP pressure probe array and the mean of the BOTTOM pressure probe array (equation \ref{VertGrad}), while $\partial C_P / \partial z$ is based on the difference between the average of the LEFT pressure probe array and the mean of the RIGHT pressure probe array (equation \ref{HorizGrad}). $\Delta L_y$ is the separation distance between the horizontally oriented pressure probe arrays and $\Delta L_z$ is the separation distance between the vertically oriented pressure probe arrays.
\begin{equation}
    \frac{ \partial C_P}{ \partial y} = H \frac{ \bar{C}_{P_T} - \bar{C}_{P_B}}{ \Delta L_y}
    \label{VertGrad}
\end{equation}
\begin{equation}
    \frac{ \partial C_P}{ \partial z} = H \frac{ \bar{C}_{P_L} - \bar{C}_{P_R}}{ \Delta L_z}
    \label{HorizGrad}
\end{equation}
Furthermore, vortex structures are identified using the $Q$-criterion. This quantity is defined as the Laplacian of pressure, $P$, which can be re-written to give the difference between the double contraction of, respectively, the rotation tensor, $\textbf{W}$, and the strain rate tensor, $\textbf{S}$ \cite{RefWorks:283}. As shown in equation \ref{DimQCriterion}, rotation dominates for positive values of $Q$, illustrating the $Q$-criterion's suitability for identifying vortex structures. The $Q$-criterion is non-dimensionalized using the body base height, $H$, and the free-stream velocity's magnitude, $|\bm{U}_{\infty}|$, to give $Q^*$ (equation \ref{NonDimQCriterion}).
\begin{equation}
    Q= \nabla^2 P =  \textbf{W} \cdot \cdot \textbf{W} - \textbf{S} \cdot \cdot \textbf{S}
    \label{DimQCriterion}
\end{equation}
\begin{equation}
    Q \quad \rightarrow \quad Q^* = \frac{Q H^2}{|\bm{U}_{\infty}|^2}
    \label{NonDimQCriterion}
\end{equation}

\subsection{Mesh construction} \label{MeshRefineStudy}
\begin{figure}
\centering
\graphicspath{ {ComputationalSetUp/} }
\includegraphics[scale=0.41]{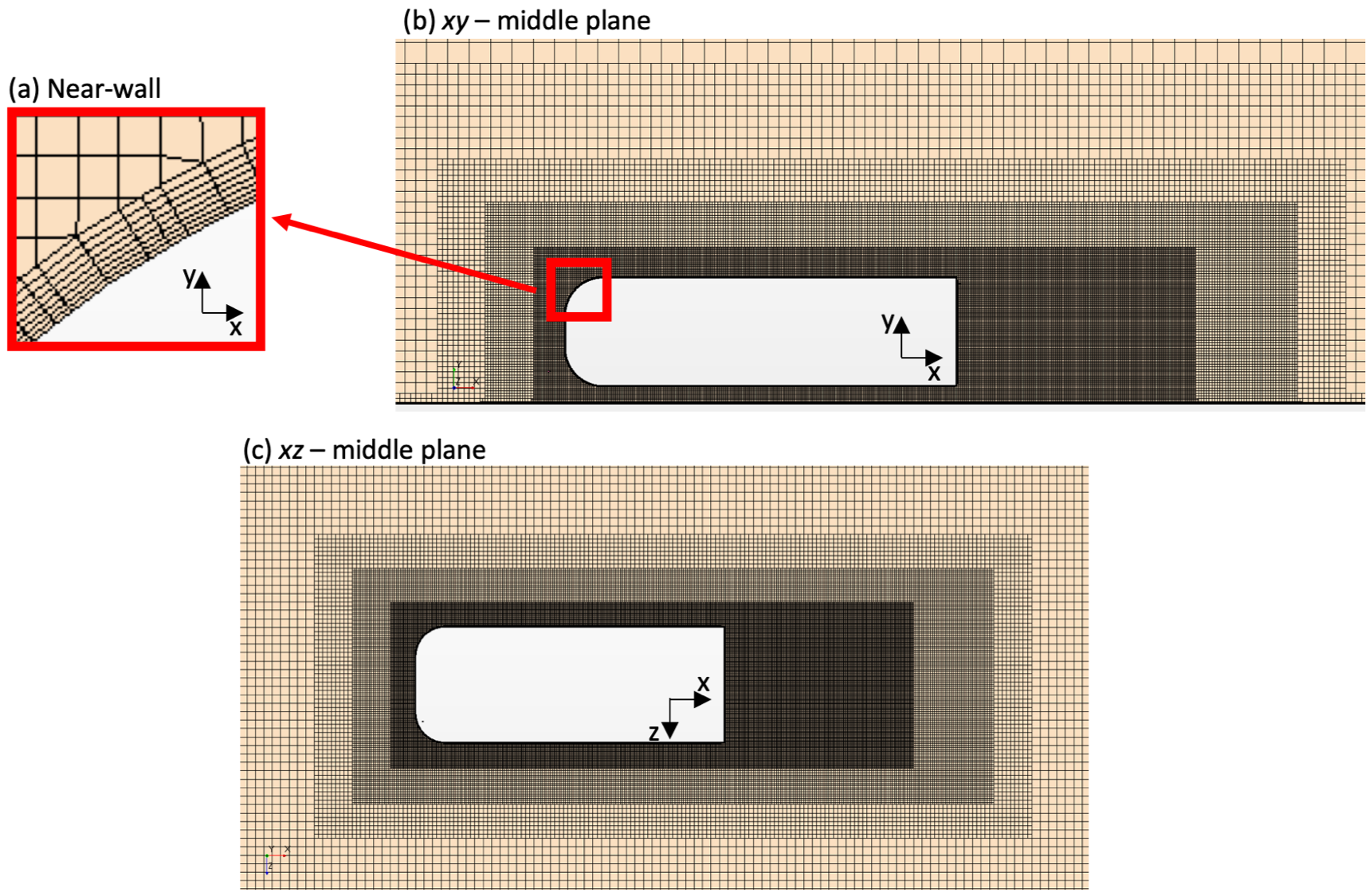}
\caption{Example volume mesh with (a) prism layer cells near the Ahmed body nose boundary, and trimmer cells in the majority of the flow domain visualized on the (b) $xy$ - middle plane and (c) $xz$ - middle plane}
\label{fig:SampleMesh}
\end{figure}

The meshes for this investigation were constructed using Star-CCM+'s built-in meshing functionality, after which they were converted to OpenFOAM format using a mesh conversion utility. Most of the flow domain was discretized using trimmer cells (figure \ref{fig:SampleMesh}(b) and \ref{fig:SampleMesh}(c)). These are hexahedrals cells that on approach to a boundary get \enquote{trimmed} to form polyhedral cells, thereby conforming to the geometry's shape. This cell type was chosen, as hexahedral cells are computationally efficient for flows with a primary direction. On approach to wall boundaries (i.e. at the ground and the Ahmed body), prism layer cells are constructed (figure \ref{fig:SampleMesh}(a)). These are orthogonal cells that are well-suited to resolving the boundary layer's large velocity gradient without exorbitantly increasing the cell count in the far-field of the domain. The resulting mesh consisting of trimmer cells and prism layer cells is an unstructured mesh. 

\begin{table}[H]
  \begin{center}
  \begin{tabular}{lccccc}
  \hline
      Mesh size          & $\Delta L_{wake}$  &  $\bar{C}_D$   & $\bar{C}_L$ & $\bar{C}_{PB}$ & $\bar{L}_B$ \\[3pt]
  \hline
      5.4 million         & 0.825$\lambda_T$  & 0.375  & -0.0542  & -0.203 & 1.417H \\
      11.1 million        & 0.413$\lambda_T$  & 0.364  & -0.0548  & -0.201 & 1.417H \\
      25.6 million        & 0.205$\lambda_T$  & 0.376  & -0.0564  & -0.214 & 1.375H \\
      Grandemange et al. \cite{RefWorks:201} & - & - & - & -0.190  & 1.418H \\
  \hline
  \end{tabular}
  \caption{Mesh refinement study for WRLES - mean flow quantities}
  \label{tab:MeshRefineStudyMeanQ}
  \end{center}
\end{table}
\begin{figure}
\centering
\graphicspath{ {ComputationalSetUp/} }
\includegraphics[scale=0.32]{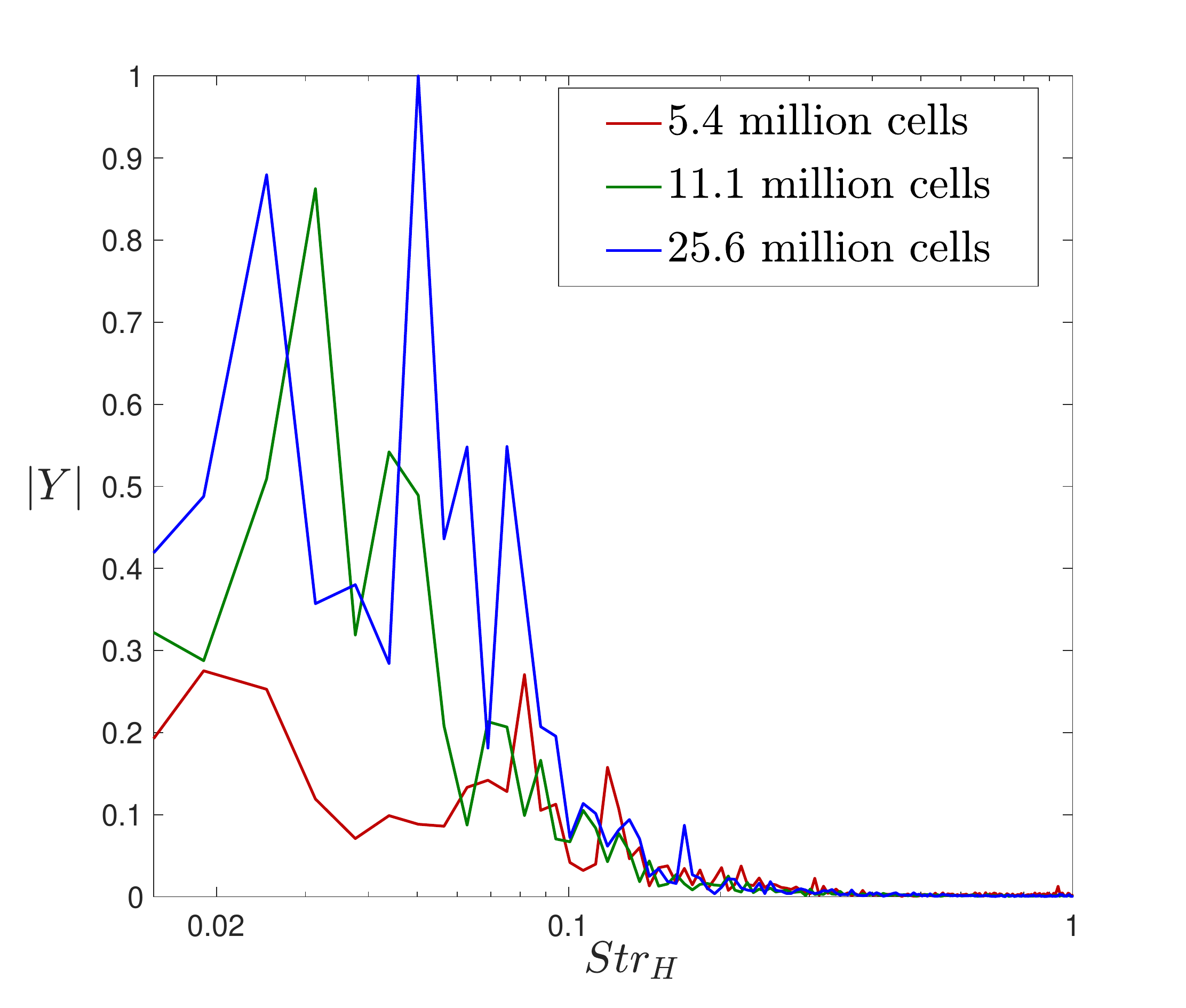}
\caption{Mesh refinement study - base pressure power spectra}
\label{fig:MeshRefineStudyPS}
\end{figure}

A mesh refinement study using WRLES was performed to identify the baseline free-stream and wake mesh for the entire study. To obtain an estimate of the wake cell size, $\Delta L_{wake}$, needed to resolve the Taylor microscale, $\lambda_T$, which is the smallest scale that LES can capture, scaling equation \ref{TaylorScale} for $\lambda_T$ was used. This relationship is derived under the assumption of a steady, homogeneous, and pure shear flow in which production of turbulent energy is equal to its viscous dissipation. Constant $A$ is originally an undetermined constant assumed to be of order one, while $l$ is the integral (large) eddy length-scale \cite{RefWorks:319}. As was done in previous Ahmed body simulation work, $A$ is taken to be equal to 0.5 and $l$ is equal to the Ahmed body height, $H$ \cite{RefWorks:321,RefWorks:320}. 
\begin{equation}
    \lambda_T \sim 15^{\frac{1}{2}} A^{-\frac{1}{2}} Re_l^{-\frac{1}{2}} l \quad \rightarrow \quad \lambda_T \sim 5.5 Re_H^{-\frac{1}{2}} H
    \label{TaylorScale}
\end{equation}
A total of three meshes were employed in the mesh refinement study. Their cell counts are 5.4 million, 11.1 million and 25.6 million. The cell size in the far-field is identical for all three meshes. Only their wake cell size differs, which for the 5.4 million, 11.1 million and 25.6 million cell mesh corresponds to 0.825$\lambda_T$, 0.413$\lambda_T$ and 0.205$\lambda_T$, respectively. 

Table \ref{tab:MeshRefineStudyMeanQ} compares the mean drag coefficient, $\bar{C}_D$, lift coefficient, $\bar{C}_L$, base pressure coefficient, $\bar{C}_{PB}$, and re-circulation bubble length, $\bar{L}_B$, for the three meshes. The values compare well between the three set-ups. A comparison to the $\bar{C}_{PB}$ and $\bar{L}_B$ obtained in the experimental study from Grandemange et al. \cite{RefWorks:201}, who looked at the squareback Ahmed body at $Re_H$=33,333, is favourable for all three meshes as well. Based on mean flow quantities, it seems reasonable to use the coarsest mesh, 5.4 million cells. 

\begin{figure}
\centering
\graphicspath{ {ComputationalSetUp/} }
\includegraphics[scale=0.23]{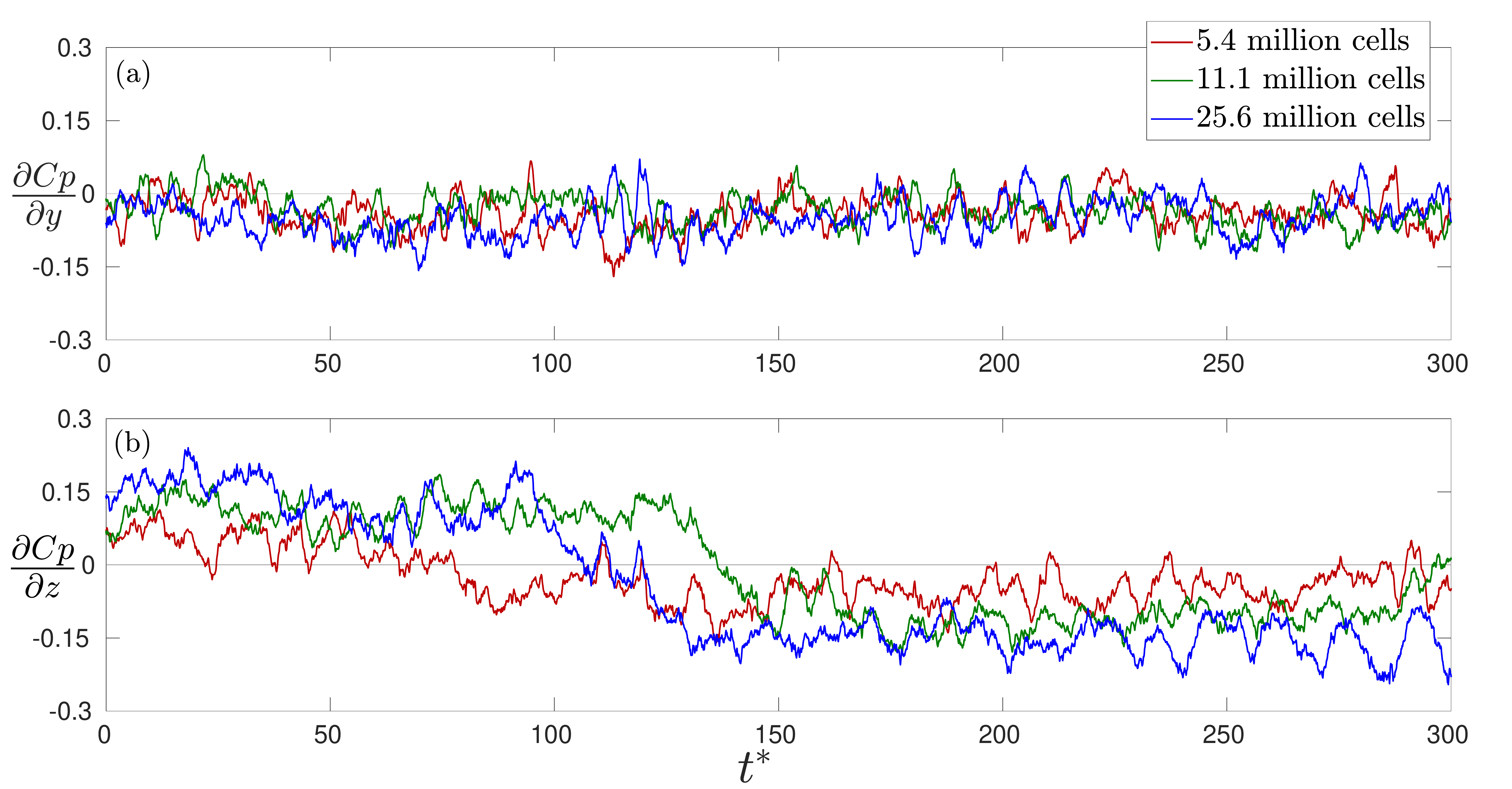}
\caption{Mesh refinement study - base pressure gradients in the (a) vertical direction, $\partial C_P / \partial y$, and (b) horizontal direction, $\partial C_P / \partial z$}
\label{fig:MeshRefineStudyGradient}
\end{figure}

However, by looking at the instantaneous flow it becomes clear that a finer mesh is needed. Figure \ref{fig:MeshRefineStudyPS} depicts the base pressure power spectra and figure \ref{fig:MeshRefineStudyGradient} presents the base pressure gradients over 300 non-dimensional time units, $t^{*}=t|\bm{U}_{\infty}|/H$, for the three meshes. The energy content of the average base pressure is significantly under-predicted with the 5.4 million cell mesh (figure \ref{fig:MeshRefineStudyPS}). There is a far stronger energy content in the average base pressure signal for the 11.1 million cell and 25.6 million cell mesh at Strouhal numbers based on body height, $Str_H$, of \textit{O}(0.01). Also, although there is no discernible difference in vertical base pressure gradient, $\partial C_P / \partial y$, for the three set-ups (figure \ref{fig:MeshRefineStudyGradient}(a)), the same is not true of the horizontal base pressure gradient, $\partial C_P / \partial z$, where the magnitude of the gradient for the 5.4 million cell mesh is noticeably smaller than for the 11.1 million and 25.6 million cell mesh (figure \ref{fig:MeshRefineStudyGradient}(b)). It was, therefore, deemed necessary to adopt the finer 11.1 million cell mesh for the WRLES simulation performed in this study. Additionally, it is noteworthy that all three WRLES meshes exhibit a bimodal event in the presented time-span of figure \ref{fig:MeshRefineStudyGradient}: the 5.4 million cell mesh at $t^{*}\sim70$, the 11.1 million cell mesh at $t^{*}\sim135$ and the 25.6 million cell mesh at $t^{*}\sim110$. This illustrates that, as part of the mesh refinement study, an objective of the investigation has already been achieved: the creation of a practical numerical set-up that features bimodality for future LES based drag reduction work. Specifically, the 11.1 million cell WRLES mesh adopted for this study is 80\% smaller than the fully block-structured WRLES meshes used in the only other published computational investigation on wake bimodality, which was performed by Dalla Longa et al. \cite{RefWorks:306}.  

\begin{figure}[H]
\centering
\graphicspath{ {ComputationalSetUp/} }
\includegraphics[scale=0.7]{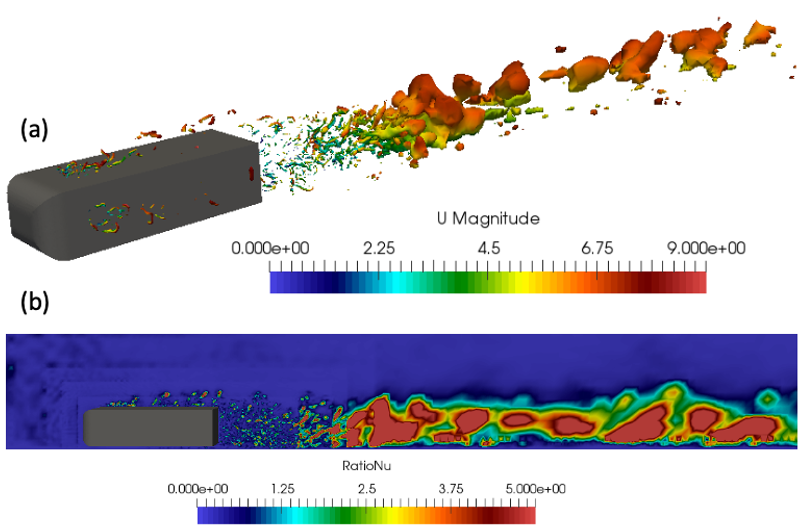}
\caption{Visualization of viscosity ratio with an (a) iso-contour plot at $\nu_t/\nu = 5$ colored by velocity magnitude and (b) a viscosity ratio contour plot}
\label{fig:NuRatio}
\end{figure}

\begin{figure}[H]
\centering
\graphicspath{ {ComputationalSetUp/} }
\includegraphics[scale=0.4]{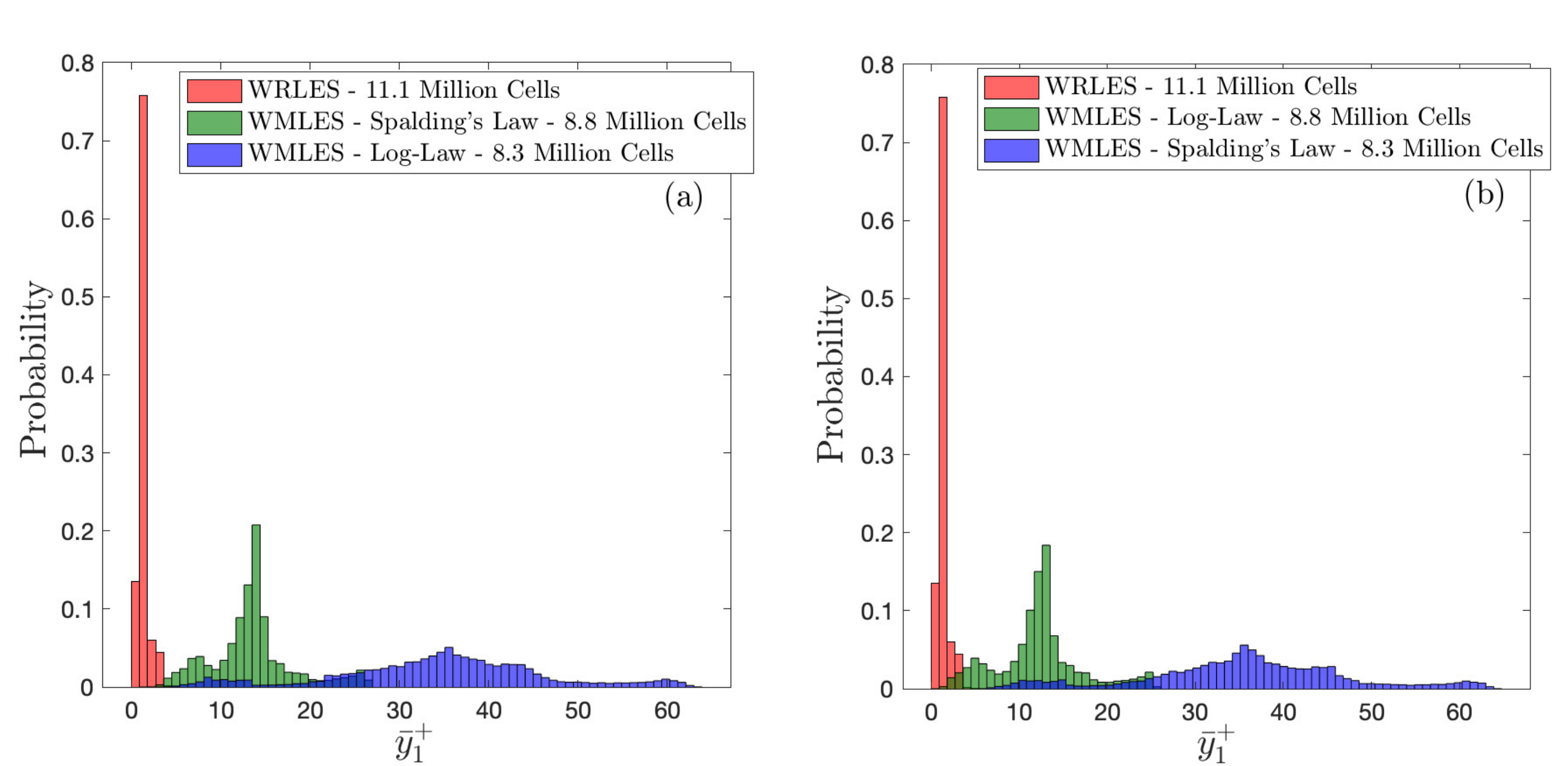}
\caption{Histogram plots of $\bar{y}_1^{+}$}
\label{fig:WallYPlusHistogram}
\end{figure}

\begin{figure}[H]
\centering
\graphicspath{ {ComputationalSetUp/} }
\includegraphics[scale=0.38]{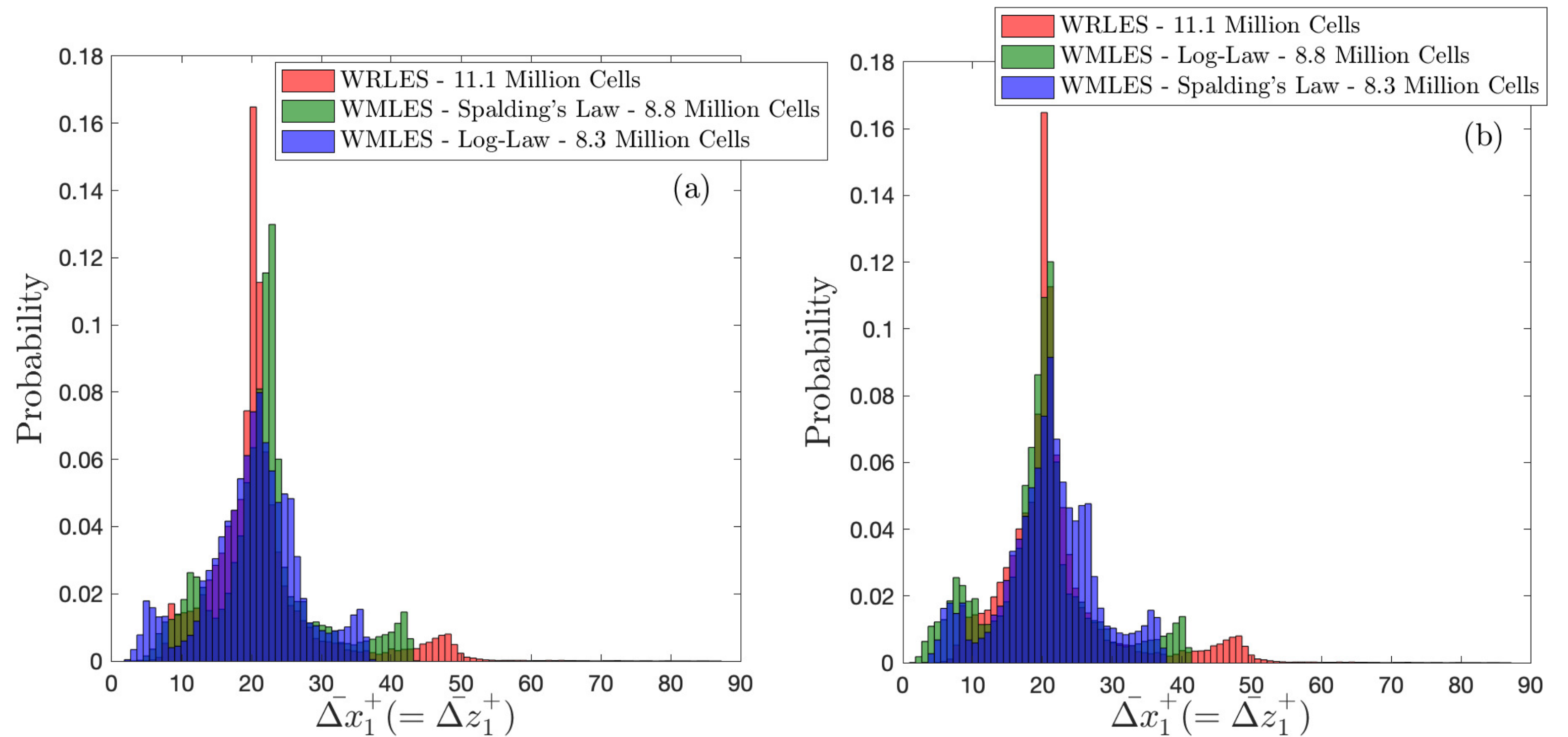}
\caption{Histogram plots of $\bar{\Delta x}_1^+$ and $\bar{\Delta z}_1^+$}
\label{fig:WallZXPlusHistogram}
\end{figure}

\begin{figure}[H]
\centering
\graphicspath{ {ComputationalSetUp/} }
\includegraphics[scale=0.43]{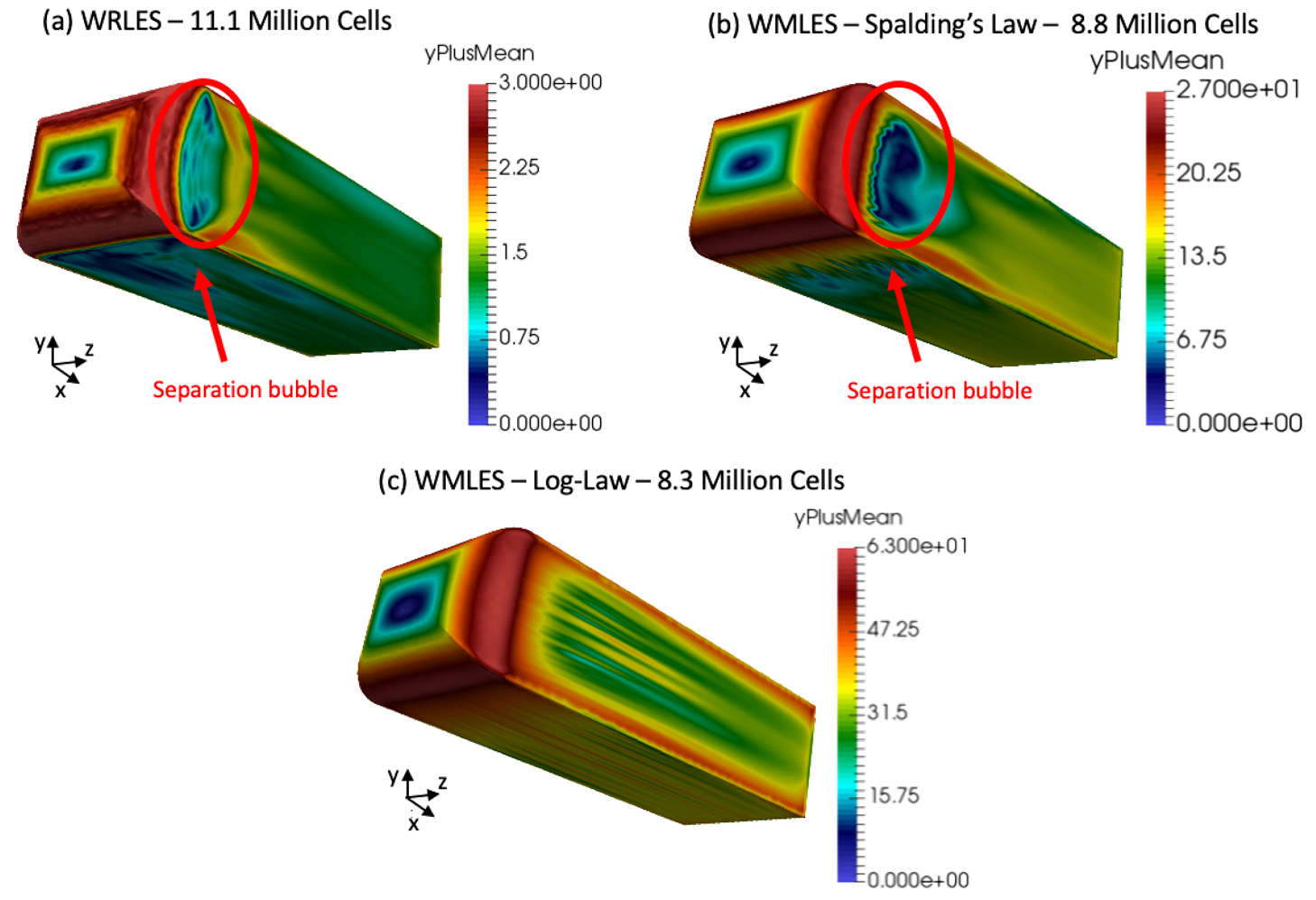}
\caption{$\bar{y}_1^{+}$ distribution projected onto the Ahmed body for (a) WRLES set-up with 11.1 million cells, (b) WMLES set-up using Spalding's law with 8.8 million cells and (c) WMLES set-up using the log-law with 8.3 million cells }
\label{fig:WallYPlusAhmedBody}
\end{figure}

\begin{figure}[H]
\centering
\graphicspath{ {ComputationalSetUp/} }
\includegraphics[scale=0.3]{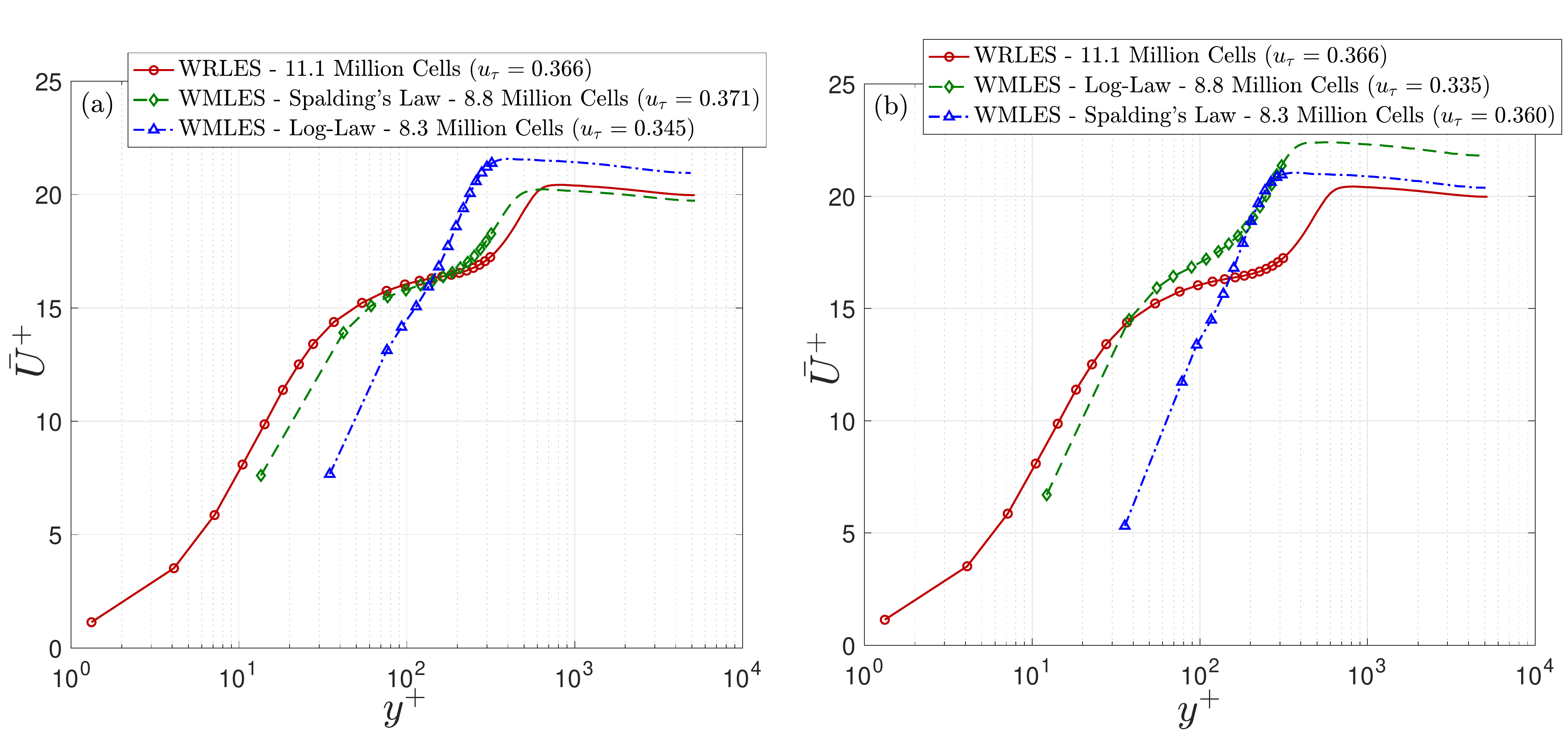}
\caption{Average turbulent boundary layer profiles in wall units taken from the Ahmed body cross-stream side(s) downstream of the front separation bubble}
\label{fig:UPlusYPlus}
\end{figure}

Added confidence in the chosen 11.1 million cell WRLES mesh is provided through its favourable comparison to further general LES mesh resolution criteria both away from and near the wall. For one, the mesh away from the wall fulfills the LES viscosity ratio criterion, $\nu_t/\nu \sim \leq 5$. As the instantaneous iso-contour plot of $\nu_t/\nu = 5$ (figure \ref{fig:NuRatio}(a)) and the accompanying viscosity ratio contour plot (figure \ref{fig:NuRatio}(b)) show, $\nu_t/\nu$ is merely greater than 5 in eddy cores closer to the Ahmed body and in the far-field wake region. Important here is to distinguish between the different causes for these elevated viscosity ratio values. As Howard and Pourquie \cite{RefWorks:320} state in their LES study of the Ahmed reference model, there are the physical strain effects near the Ahmed body wall and then there are the artificial grid effects in the wake’s far-field region. The large cell size in the far-field wake amplifies even negligible strain effects due to the WALE turbulence model’s grid dependent length-scale for the eddy viscosity. Therefore, the raised viscosity ratio values in the far-field wake may be ignored. From here on, the wake and far-field mesh remain unchanged from the above mentioned baseline WRLES mesh. Subsequently constructed WMLES meshes have the same wake and far-field mesh construction. Moving to the 11.1 million cell WRLES mesh's near-wall topology, it has 8 prism layer cells (figure \ref{fig:SampleMesh}(a)), as the mesh must be able to fully resolve the large velocity gradient of the turbulent boundary layer. Its average wall y-plus, $\bar{y}_1^{+}$, distribution must be centered around 1. This is the case, as illustrated in the histogram plots of figure \ref{fig:WallYPlusHistogram}. By projecting the $\bar{y}_1^{+}$ distribution onto the Ahmed body, it is evident that $\bar{y}_1^{+}>1$ values are located around the Ahmed body nose (figure \ref{fig:WallYPlusAhmedBody}(a)). This near-wall wall-normal mesh resolution is deemed acceptable, as the flow is likely still laminar around the nose. Transition to turbulence is expected to occur as a result of flow separation just aft of the Ahmed body nose, where a separation bubble forms (see sub-section \ref{SourceWakeBimodal}). In fact, traces of this separation bubble can be seen in the $\bar{y}_1^{+}$ distribution (figure \ref{fig:WallYPlusAhmedBody}(a)). Further near-wall WRLES mesh criteria include average streamwise cell spacing, $\bar{\Delta x}_1^+$, of 50-130, and cross-stream grid spacing, $\bar{\Delta z}_1^+$, of 15-30, as well as 10-30 grid-points populated across the inner region of the turbulent boundary layer ($\bar{y}_1^{+}<100$) \cite{LES_Mesh}. For meshes in this work, $\bar{\Delta x}_1^+$ is equal to $\bar{\Delta z}_1^+$ i.e. the cells are isotropic in the streamwise and cross-stream direction. As histogram plots in figure \ref{fig:WallZXPlusHistogram} show, the WRLES grid is characterized by $\bar{\Delta x}_1^+$ and $\bar{\Delta z}_1^+$ values centered about 20. This means the WRLES grid is slightly over-resolved in the streamwise direction, but fulfills the grid criteria in the cross-stream direction as stipulated by Choi and Moin \cite{LES_Mesh}. Figure \ref{fig:UPlusYPlus} also shows that 11 cells are positioned within the inner region of the turbulent boundary layer, thereby fulfilling the final near-wall mesh criterion.

Having conclusively identified the baseline WRLES mesh of 11.1 million cells, two meshes for WMLES simulations were derived from it. With the mesh away from the wall fixed, only the near-wall prism layer mesh was altered. The first WMLES mesh created replaces the 8 prism layers with just 2 prism layers. In total, the mesh has 8.8 million cells, a reduction of about 20\% in cell count from the baseline 11.1 million cell WRLES mesh. Both wall-models, Spalding's law (equation \ref{SpaldLaw}) and the log-law (equation \ref{LogLaw}), were tested with this mesh. The average wall y-plus, $\bar{y}_1^{+}$, distribution for both 8.8 million cell wall-model simulations is illustrated in figure \ref{fig:WallYPlusHistogram} for (a) Spalding's law and (b) the log-law. It seems the wall-models give very similar $\bar{y}_1^{+}$ distributions suggesting that the wall-model used has relatively little impact on the flow resolved. The histogram plot centers are positioned well beyond $\bar{y}_1^{+}=10$, which is in line with the lower resolution requirements associated with wall-models. Furthermore, projecting the $\bar{y}_1^{+}$ distribution onto the Ahmed body (figure \ref{fig:WallYPlusAhmedBody}(b) for Spalding's law) returns a figure similar to the WRLES case (figure \ref{fig:WallYPlusAhmedBody}(a)), particularly the trace of the front separation bubble. Piomelli and Balaras \cite{RefWorks:231} further stipulate that $\bar{\Delta x}_1^+ \sim 1500$ and $\bar{\Delta z}_1^+ \sim 70$ in order for the Reynolds-averaging effect to be sufficient for wall-model use. These requirements are not met by the constructed WMLES mesh, as altering the grid in the streamwise and spanwise direction would influence the mesh away from the wall throughout the domain, thereby altering the far-field mesh. Thus, the $\bar{\Delta x}_1^+$ and $\bar{\Delta z}_1^+$ values for the 8.8 million cell WMLES mesh continue to be centered about 20, as is the case for the WRLES mesh (figure \ref{fig:WallZXPlusHistogram}). It is nonetheless appropriate to call these simulations WMLES, as the wall-normal grid resolution has been reduced to a suitable value and, ultimately, the no-slip condition is modelled. The second WMLES mesh involves a further reduction of the prism layers used. In the final WMLES mesh, only 1 prism layer is used. This results in an 8.3 million cell mesh, which too was tested with both wall-models, (a) the log-law and (b) Spalding's law (figure \ref{fig:WallYPlusHistogram}). Their $\bar{y}_1^{+}$ distributions are both centered beyond 30 and, again, show little deviation from each other. This coarser near-wall mesh is associated with more Reynolds-averaging of wall-normal fluctuations, which translates to a noticeable difference in flow structures imprinted onto the Ahmed body $\bar{y}_1^{+}$ distribution. For example, in figure \ref{fig:WallYPlusAhmedBody}(c) for the log-law, no front separation bubble is present and this plays a prominent role throughout section \ref{SourceWakeBimodal}. 

\subsection{Wall-model analysis} \label{WMAnalysis}
Based on the presented data for $\bar{y}_1^{+}$, $\bar{\Delta x}_1^+$ and $\bar{\Delta z}_1^+$, it seems plausible that both wall-models can be used with the 8.8 and 8.3 million cell WMLES meshes. However, there are differences and these become apparent when looking at averaged boundary layer profiles in wall units taken from the Ahmed body cross-stream side downstream of the separation bubble where the flow has recovered (figure \ref{fig:UPlusYPlus}). It seems that for the 8.8 million cell mesh, better correlation with the WRLES curve is attained when using Spalding's law. This is in line with the observation that the log-law should ideally be used for set-ups with the wall-adjacent cell center beyond the buffer region of the turbulent boundary layer, a criterion that the 8.8 million cell mesh does not fulfill. Nonetheless, the 8.8 million cell mesh with the log-law does still better capture the energetic buffer region of the turbulent boundary layer than either of the two simulations performed with the 8.3 million cell mesh. The 8.3 million cell mesh's inability to correctly resolve the transition from the buffer region to the fully turbulent log region of the boundary layer with either wall-model means that important physics is not captured, physics that, as shown in \ref{SourceWakeBimodal}, is needed for wake bimodality. The remainder of this work focuses primarily on three simulations: the 11.1 million cell mesh with WRLES, the 8.8 million cell mesh with Spalding's law of the wall and the 8.3 million cell mesh with the log-law of the wall. References are also made to complementing material found in an Appendix.  

\section{Results} \label{Results}
This section on the study's key findings is divided into two subsections: detecting wake bimodality and identifying a likely trigger of wake bimodality. The goal is to first determine which of the three simulations of varying near-wall resolution (i.e. the WRLES mesh and the two WMLES mesh constructions) captures bimodality and, subsequently, to explain why a given set-up is able to resolve a bimodal event based on the large-scale coherent flow structures present in the simulation. 

Prior to the analysis of bimodality, it is apt to look at the mean aerodynamic quantities for the WMLES simulations (table \ref{tab:WRWMLESMeanQ}). These are positioned alongside the corresponding baseline WRLES simulation values. Taking the WRLES data as a benchmark, there is some variability among the displayed quantities: up to 5\% for drag, 30\% for lift, 22\% for base pressure and 6\% for the wake length. It is known that in aerodynamic studies mean lift is associated with a larger error margin than drag and as such the significant variability reported for lift is not entirely unexpected. More importantly, it may be observed that deviation is particularly strong for $\bar{C}_L$ and $\bar{C}_{PB}$ with the WMLES simulations using the 8.3 million cell mesh. This alludes to the fundamentally different flow captured in the simulation using 8.3 million cells, as discussed further in section \ref{SourceWakeBimodal}. 

\begin{table}[H]
  \begin{center}
  \begin{tabular}{lccccc}
  \hline
      Simulation          & $\Delta L_{wake}$  &  $\bar{C}_D$   & $\bar{C}_L$ & $\bar{C}_{PB}$ & $\bar{L}_B$ \\[3pt]
  \hline
      WRLES - 11.1 MC & 0.413$\lambda_T$  & 0.364  & -0.0548  & -0.201 & 1.417H\\
      WMLES - SL - 8.8 MC & 0.413$\lambda_T$ & 0.365  & -0.0506  & -0.204 & 1.403H \\
      \textit{WMLES - LL - 8.8 MC} & \textit{0.413$\lambda_T$} & \textit{0.346}  & \textit{-0.0526}  & \textit{-0.203} & \textit{1.444H} \\
      WMLES - LL - 8.3 MC & 0.413$\lambda_T$ & 0.369  & -0.0405  & -0.157 & 1.500H \\
      \textit{WMLES - SP - 8.3 MC} & \textit{0.413$\lambda_T$} & \textit{0.375}  & \textit{-0.0383}  & \textit{-0.158} & \textit{1.472H} \\
      Grandemange et al. \cite{RefWorks:201} & - & - & - & -0.190  & 1.418H \\
  \hline
  \end{tabular}
  \caption{Mean aerodynamic quantities for the WMLES simulations where MC represents "million cells", SL denotes Spalding's law and LL stands for "log-law". Complementing wall-model results are written in \textit{italics}}
  \label{tab:WRWMLESMeanQ}
  \end{center}
\end{table}

\subsection{Detection of wake bimodality} \label{DetectWakeBimodal}
Based on the work of Grandemange et al. \cite{RefWorks:201}, who looked at the presence of bimodality aft of squareback bluff-body geometries with varying aspect ratio and ground clearance, for the Ahmed body's aspect ratio and the chosen ground clearance switches in the wake's off-center position should occur in the horizontal direction. This is in the $z$-direction for the adopted coordinate system. 

\begin{figure}
\centering
\graphicspath{ {Results/} }
\includegraphics[scale=0.18]{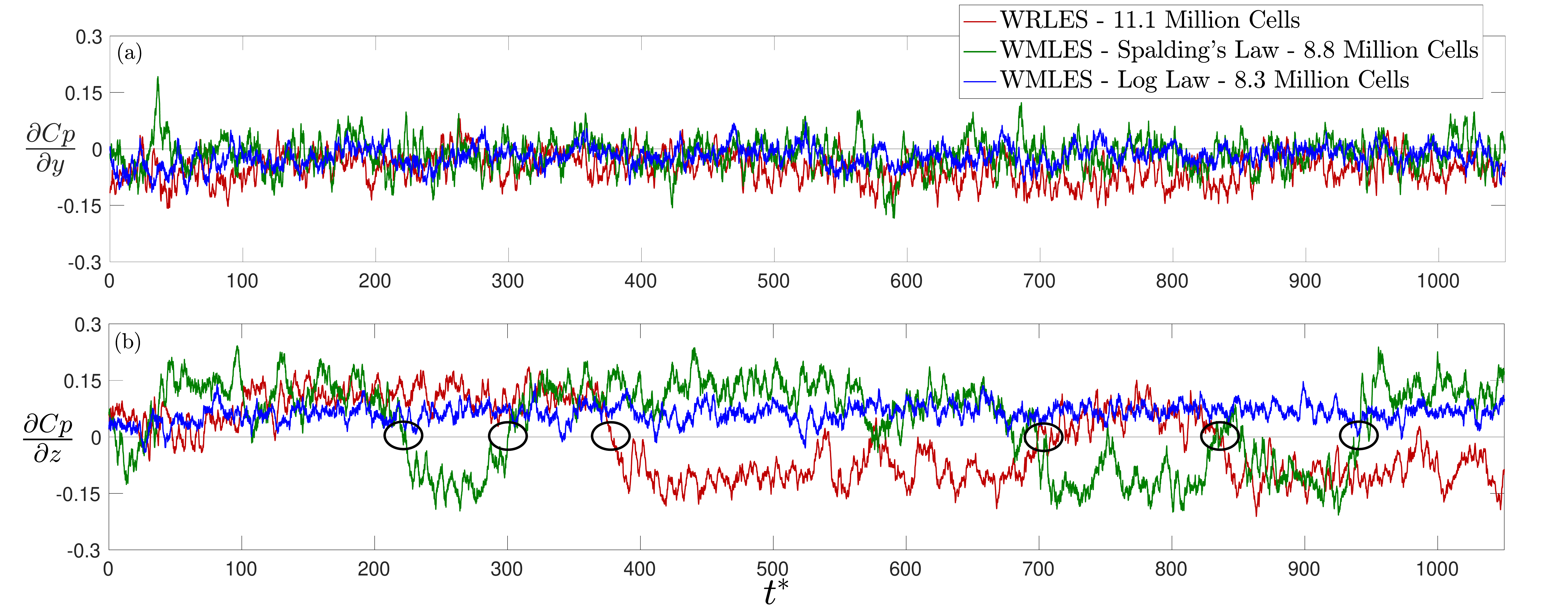}
\caption{Base pressure gradient in the (a) vertical direction, $\partial C_P / \partial y$, and (b) horizontal direction, $\partial C_P / \partial z$, for 11.1 million cells WRLES, 8.8 million cells WMLES with Spalding's law and 8.3 million cells WMLES employing the log-law. The black circles show bimodal wake shifting events}
\label{fig:BaseCPGradWRvsWMLES}
\end{figure}

\begin{figure}[ht]
\centering
\graphicspath{ {Results/} }
\includegraphics[scale=0.31]{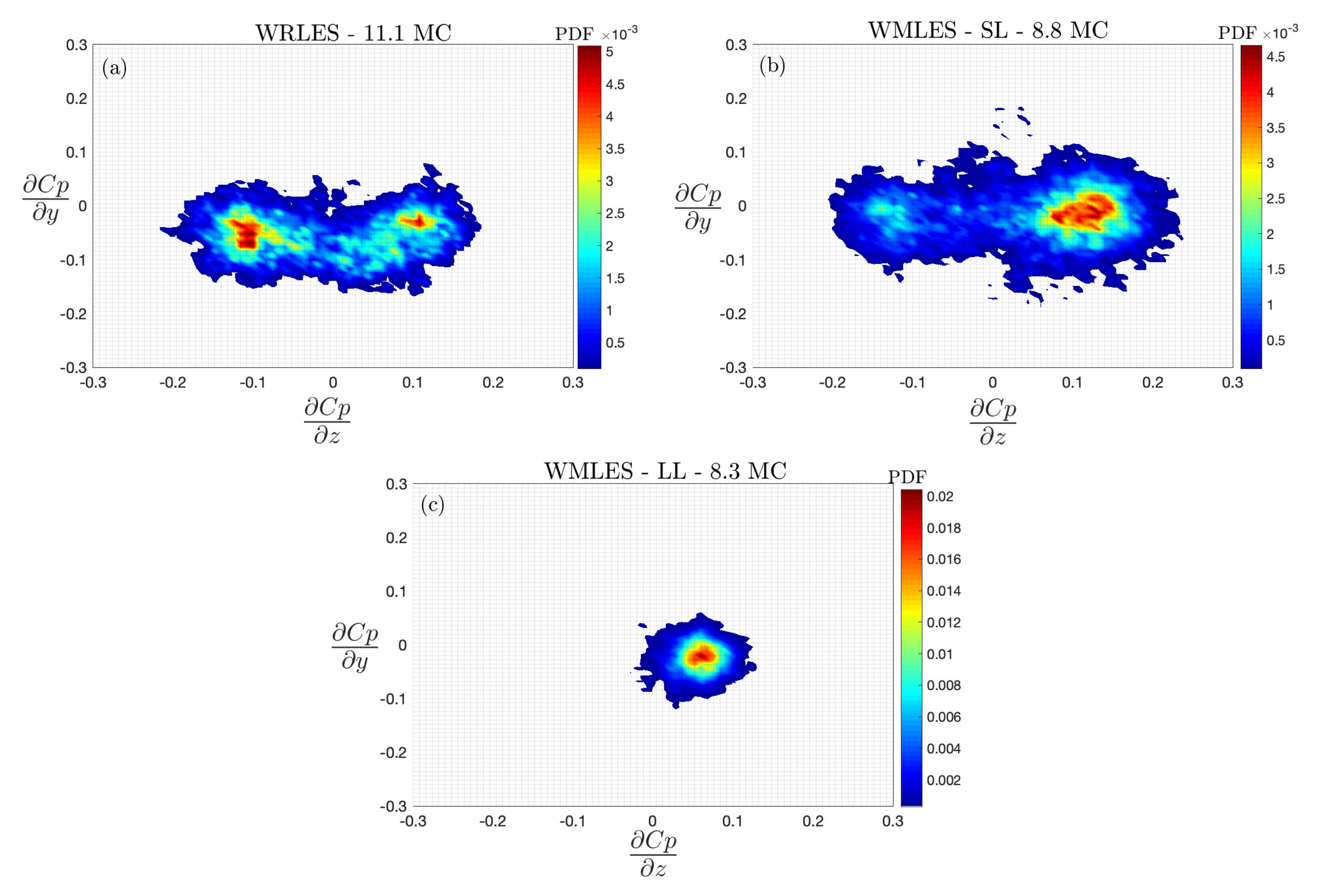}
\caption{Joint probability density function of base pressure gradients, PDF($\partial C_P/ \partial y$,$\partial C_P / \partial z$), for (a) 11.1 million cells (MC) WRLES, (b) 8.8 million cells WMLES applying Spalding's law (SL) and (c) 8.3 million cells WMLES using the log-law (LL)} 
\label{fig:JointPDF}
\end{figure}

\begin{figure}
\centering
\graphicspath{ {Appendix/} }
\includegraphics[scale=0.23]{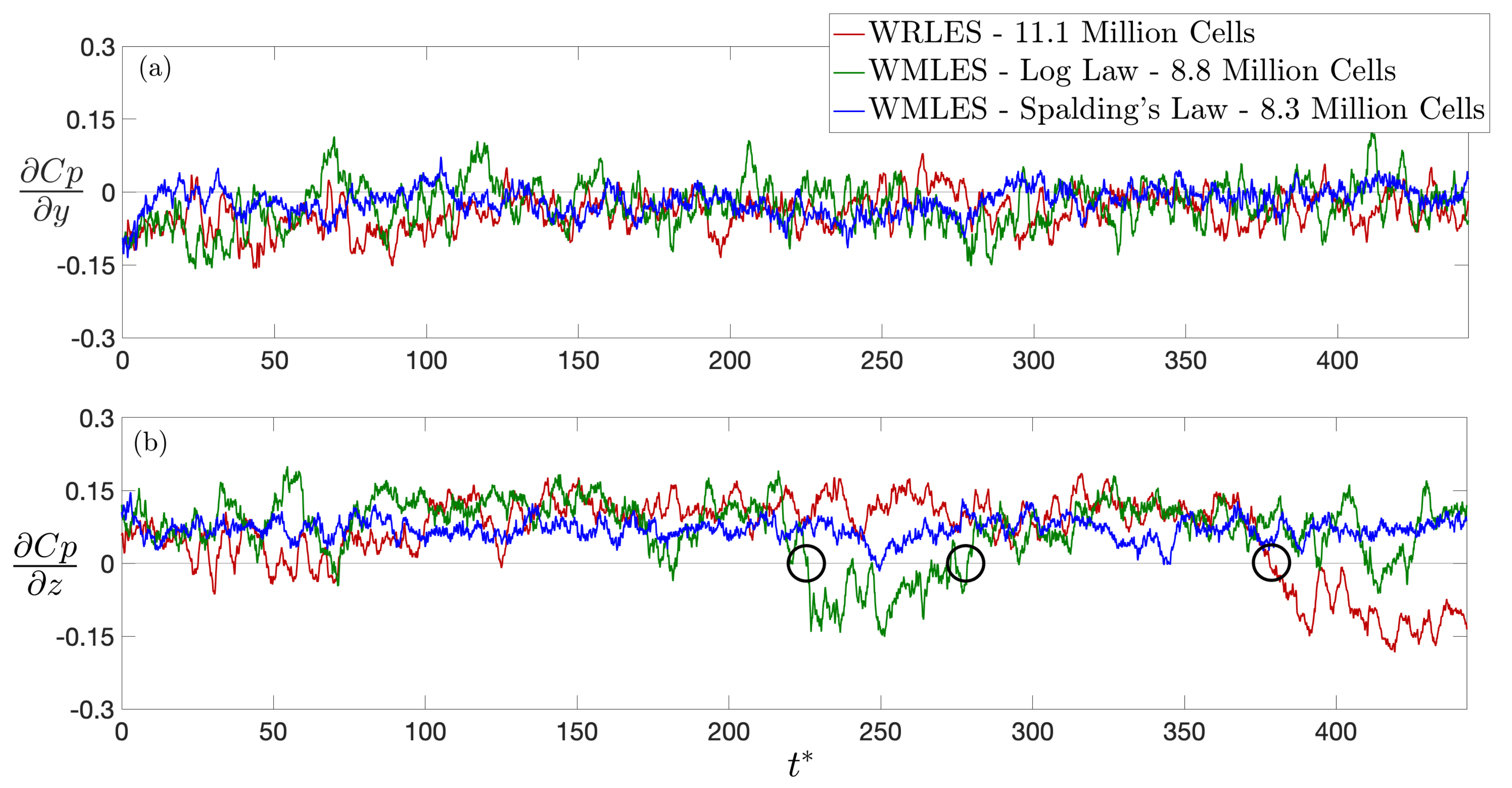}
\caption{Base pressure gradient in the (a) vertical direction, $\partial C_P / \partial y$, and (b) horizontal direction, $\partial C_P / \partial z$, for 11.1 million cells WRLES, 8.8 million cells WMLES with the log-law and 8.3 million cells WMLES employing Spalding's law. The black circles show bimodal wake shifting events}
\label{fig:BaseCPGradWRvsWMLES_Appendix}
\end{figure}

In figure \ref{fig:BaseCPGradWRvsWMLES}, base pressure gradients in the vertical direction, $\partial C_P / \partial y$, and the horizontal direction, $\partial C_P / \partial z$, are provided for the three simulation cases under closer investigation in this paper. The time spans 1050 non-dimensional time units, $t^{*}$. This time window is deemed sufficient to determine whether wake bimodality can be captured, as the characteristic time-scale over which a bimodal event occurs is up to $t^{*} \sim 10^3$ \citep{RefWorks:202}. All three simulations exhibit a mild asymmetry in the vertical direction (figure \ref{fig:BaseCPGradWRvsWMLES}(a)). Specifically, the wake is tilted toward its top trailing edge due to the Ahmed body's ground proximity, which causes the flow to accelerate in the under-body gap to the ground, dragging the wake's bottom side downstream at the gap's exit. This weak vertical asymmetry in the Ahmed body's wake is also found in the experimental work on wake bimodality by Grandemange et al. \cite{RefWorks:201} and Grandemange et al. \cite{RefWorks:202} for a $Re_H$ of 33,333 and 92,000, respectively. It can, therefore, be concluded that the wake balance between the top and bottom shear layers is accurately simulated, which, in order to capture bimodality, must be correctly resolved. This was illustrated in the experimental work of Perry et al. \cite{RefWorks:216}. They lowered the tendency of bimodality developing aft of the Windsor body, a simplified squareback geometry, through the application of tapers at the top and bottom trailing edges that modified the base pressure distribution and, hence, the vertical wake balance. Furthermore, the three test cases possess a distinct horizontal asymmetry of the wake (figure \ref{fig:BaseCPGradWRvsWMLES}(b)), indicating that the horizontal wake balance between the side shear layers is correctly simulated too. Only two mesh set-ups present bimodal events: the 11.1 million cell WRLES case and the 8.8 million cell WMLES simulation that uses Spalding's law (represents a further 21\% reduction in the mesh size needed to capture wake bimodality). For WRLES, the wake shifts its horizontal off-center position at $t^{*}$-values of 378, 713 and 836. Meanwhile, for 8.8 million cell WMLES with Spalding's law, bimodal events are recorded at $t^{*}$-values of approximately 222, 300, 698 and 941. Figure \ref{fig:JointPDF} plots the distributions of $\partial C_P / \partial y$ and $\partial C_P / \partial z$ as a joint probability density function, PDF($\partial C_P / \partial y$,$\partial C_P / \partial z$). In doing so, it is possible to further highlight the existence of bimodal events. The two simulations that capture bimodality possess two attractors, one centered at $\partial C_P / \partial z>$0 and one centered at $\partial C_P / \partial z<$0 (figure \ref{fig:JointPDF}(a) and \ref{fig:JointPDF}(b)). Meanwhile, the 8.3 million cell WMLES simulation that employs the log-law only has one attractor (figure \ref{fig:JointPDF}(c)), which is centered at $\partial C_P / \partial z>$0, illustrating the mesh's deficiency in capturing wake bimodality. It is expected that, as the simulations are extended and $t^{*} \rightarrow \infty$, thereby enabling more bimodal events to be captured, the two attractors present in figures \ref{fig:JointPDF}(a) and \ref{fig:JointPDF}(b), respectively, will exhibit color intensities that are more similar, indicating that the wake spends equal amounts of time in each asymmetric state. To further illustrate the mesh dependency associated with capturing bimodality, the Appendix contains a base pressure gradient plot (figure \ref{fig:BaseCPGradWRvsWMLES_Appendix}) that complements figure \ref{fig:BaseCPGradWRvsWMLES}. It highlights that bimodality can too be simulated with the log-law applied to the 8.8 million cell mesh while bimodality continues to be elusive for the 8.3 million cell mesh even when Spalding's law is applied.

\begin{figure}
\centering
\graphicspath{ {Results/} }
\includegraphics[scale=0.44]{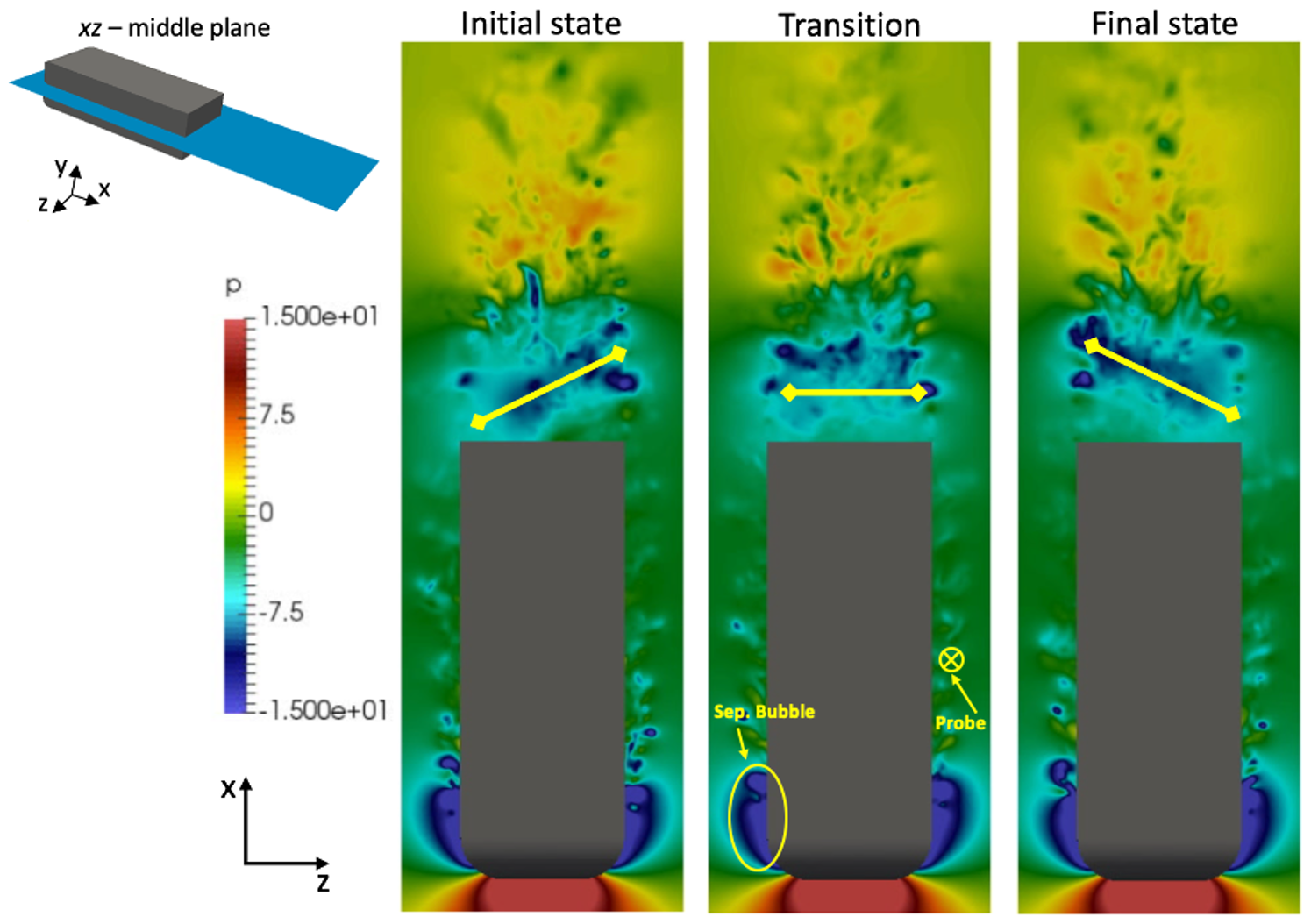}
\caption{Wake bimodality visualized using the instantaneous pressure field projected onto the $xz$ - middle plane passing through the Ahmed body for the WRLES case}
\label{fig:P_Contour_Switch}
\end{figure}

\begin{figure}[ht]
\centering
\graphicspath{ {Results/} }
\includegraphics[scale=0.44]{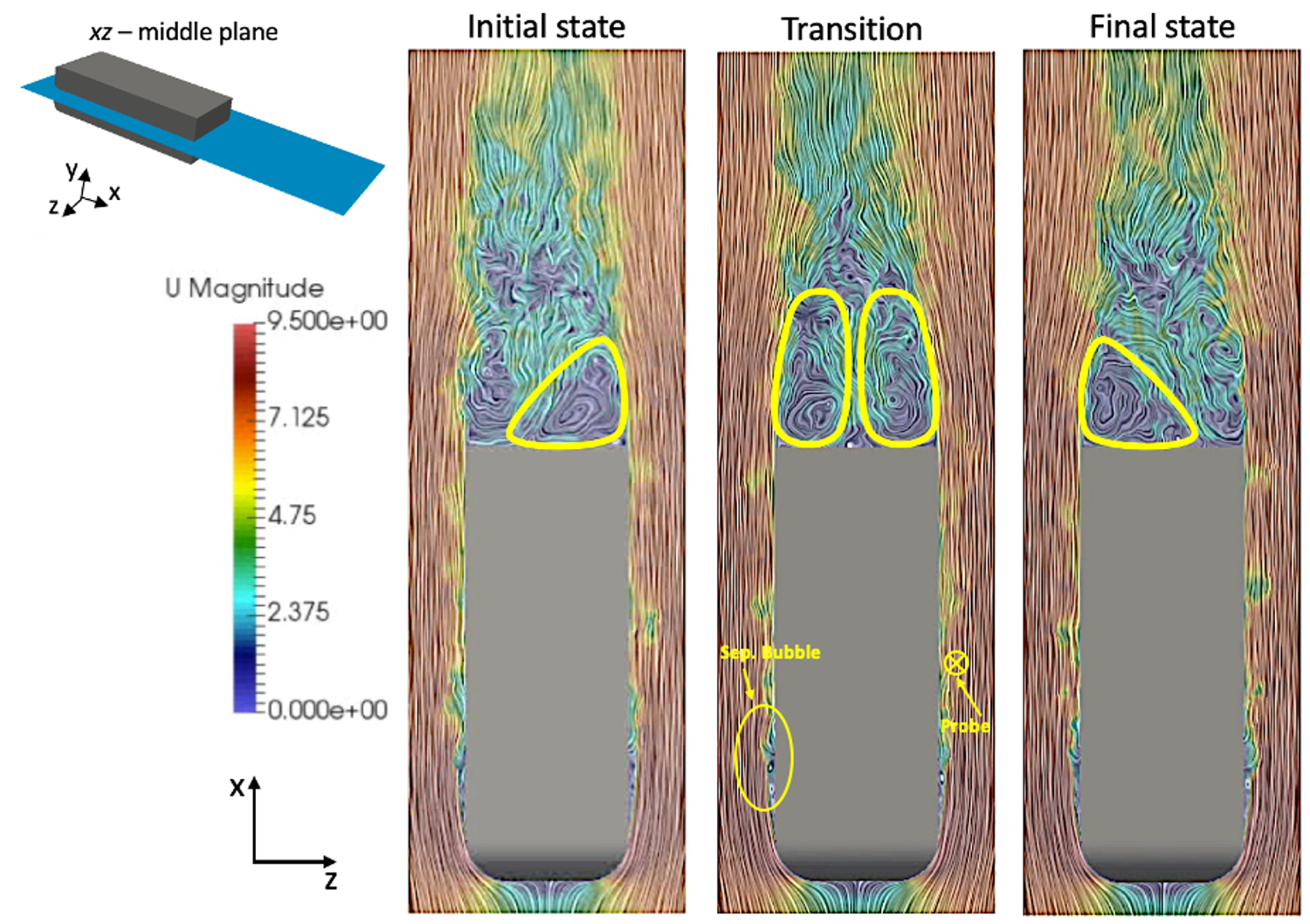}
\caption{Wake bimodality visualized using the instantaneous velocity magnitude field with streamlines projected onto the $xz$ - middle plane passing through the Ahmed body for the 8.8 million cell WMLES set-up that uses Spalding's law}
\label{fig:Umag_Contour_Switch}
\end{figure}

\begin{figure}
\centering
\graphicspath{ {Results/} }
\includegraphics[scale=0.32]{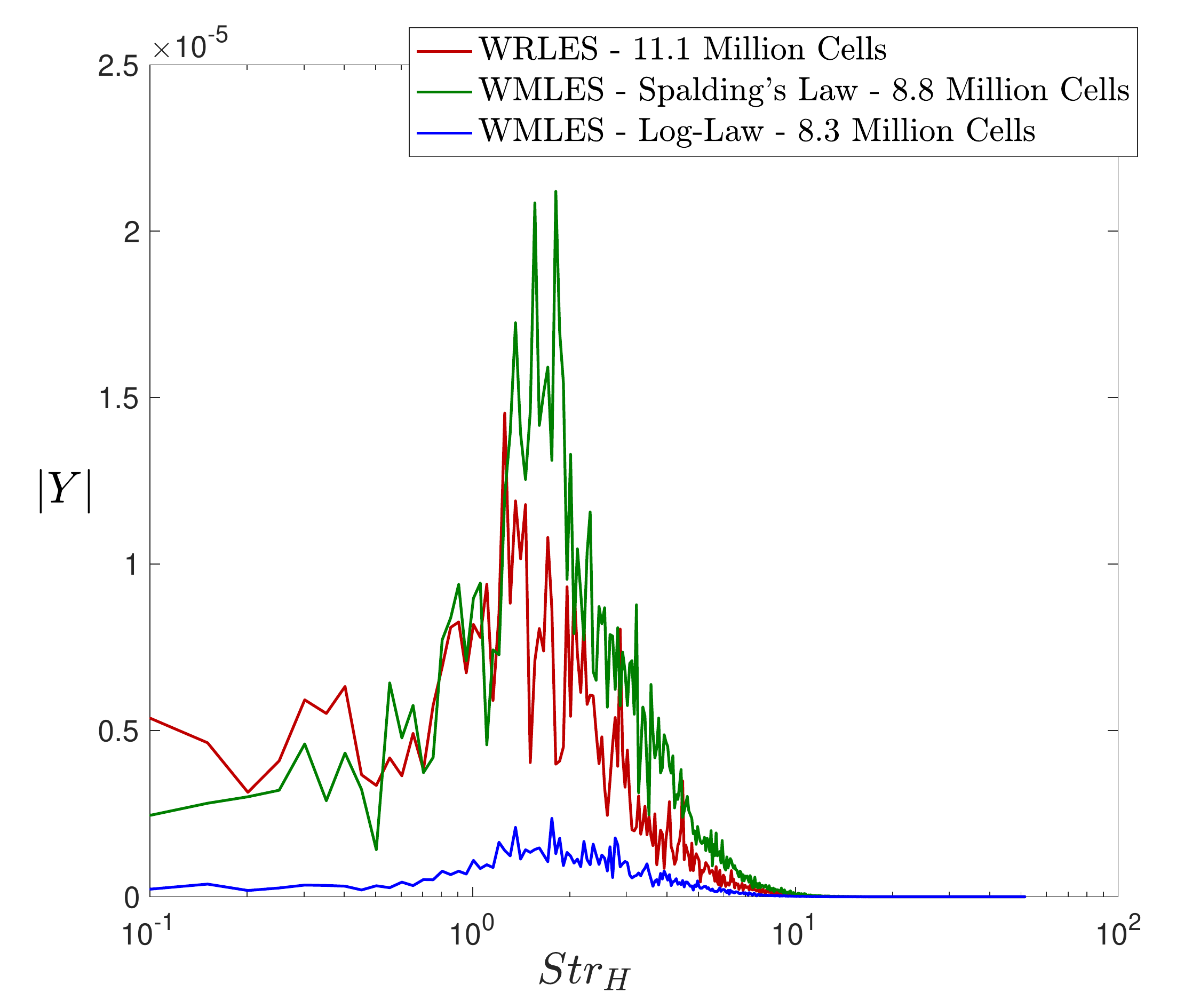}
\caption{Vortex shedding frequency associated with front separation bubble}
\label{fig:FrontSeparationBubbleFFT}
\end{figure}

\begin{figure}
\centering
\graphicspath{ {Appendix/} }
\includegraphics[scale=0.44]{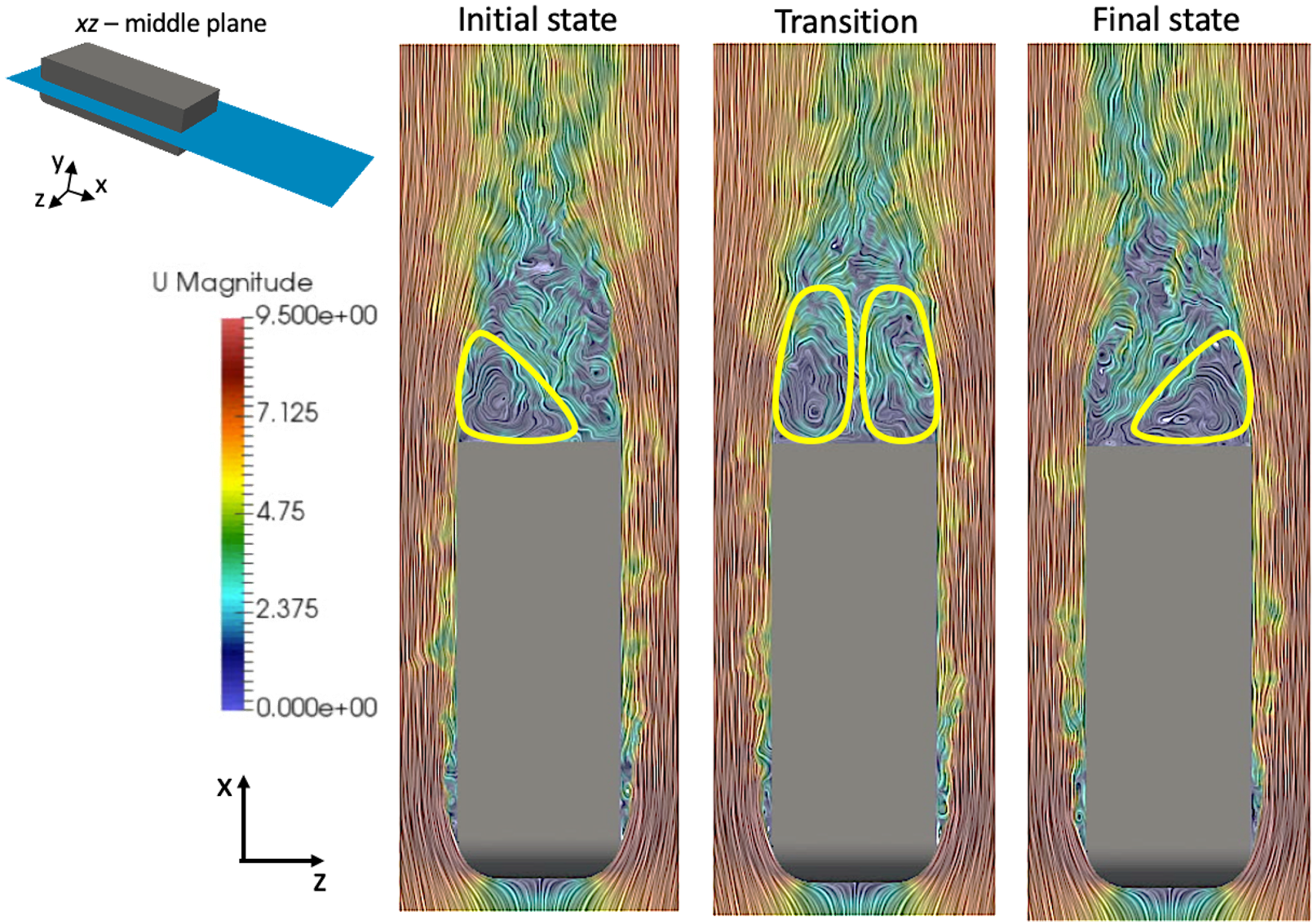}
\caption{Wake bimodality visualized using the instantaneous velocity magnitude field with streamlines projected onto the $xz$ - middle plane passing through the Ahmed body for the WRLES set-up}
\label{fig:Umag_Contour_Switch_WRLES_Appendix}
\end{figure}

Wake bimodality can also be identified by looking at pressure and velocity magnitude contour plots on the $xz$ - middle plane passing through the Ahmed body. For the 11.1 million cell WRLES simulation, instantaneous pressure contour plots for the wake tilted toward opposing off-center spanwise locations as well as a snapshot of its transitory symmetric state are presented in figure \ref{fig:P_Contour_Switch}. The wake is initially tilted to the left-hand side and shifts to the right-hand side. A low pressure separation bubble at the Ahmed body's nose is clearly identifiable as well. From this bubble, low pressure vortex structures are shed. These grow as they are convected downstream where they subsequently interact with the wake. As described in subsection \ref{SourceWakeBimodal}, it is precisely these larger scale structures that are likely to play a critical role in forcing a bimodal event. Figure \ref{fig:Umag_Contour_Switch} contains contour plots of instantaneous velocity magnitude with streamlines for the 8.8 million cell WMLES set-up that employs Spalding's law. Here, the wake is at first tilted to the right-hand side whereupon its low pressure center moves to the left-hand side. A large recirculation core is present on the spanwise side to which the asymmetric wake is tilted. As the wake transitions from one side to the other, the left and right recirculation cores become similar in size, illustrating that the wake is symmetric during this phase. The front separation bubble as well as shed vortices described previously are also identifiable in these flow field snapshots. Although different quantities are visualized in figures \ref{fig:P_Contour_Switch} and \ref{fig:Umag_Contour_Switch}, a comparison of the separation bubbles can still be made and is paramount. For the 11.1 million cell WRLES simulation the recirculation bubble is slightly more prominant in size and occurs further upstream than in the 8.8 million cell WMLES simulation with Spalding's law. This can be further visualized when looking at complementing figure \ref{fig:Umag_Contour_Switch_WRLES_Appendix} for 11.1 million cell WRLES and figure \ref{fig:P_Contour_Switch_WMLES_Appendix} for 8.8 million cell WMLES using Spalding's law in the Appendix. Spectra (figure \ref{fig:FrontSeparationBubbleFFT}) were also computed using the fast Fourier transform of pressure data recorded at a fixed position downstream of the front seperation bubble (figures \ref{fig:P_Contour_Switch} and \ref{fig:Umag_Contour_Switch}). Results indicate that, despite varying bubble size, for the chosen probe position, the shedding power of the vortices remains large for both the WRLES case and 8.8 million cell WMLES set-up using Spalding's law. Shedding amplitude is, however, largest for the latter simulation, but this should only be because the separation bubble is positioned marginally closer to the probe in this simulation set-up (figure \ref{fig:Umag_Contour_Switch}). Finally, it is important to note that the front shedding frequency is negligible for the coarse 8.3 million cell set-up that is used here with the log-law. This indicates the non-existence of the front separation bubble for the coarse wall-model simulation, a theme prevalent throughout section \ref{SourceWakeBimodal}.

\begin{figure}
\centering
\graphicspath{ {Results/} }
\includegraphics[scale=0.58]{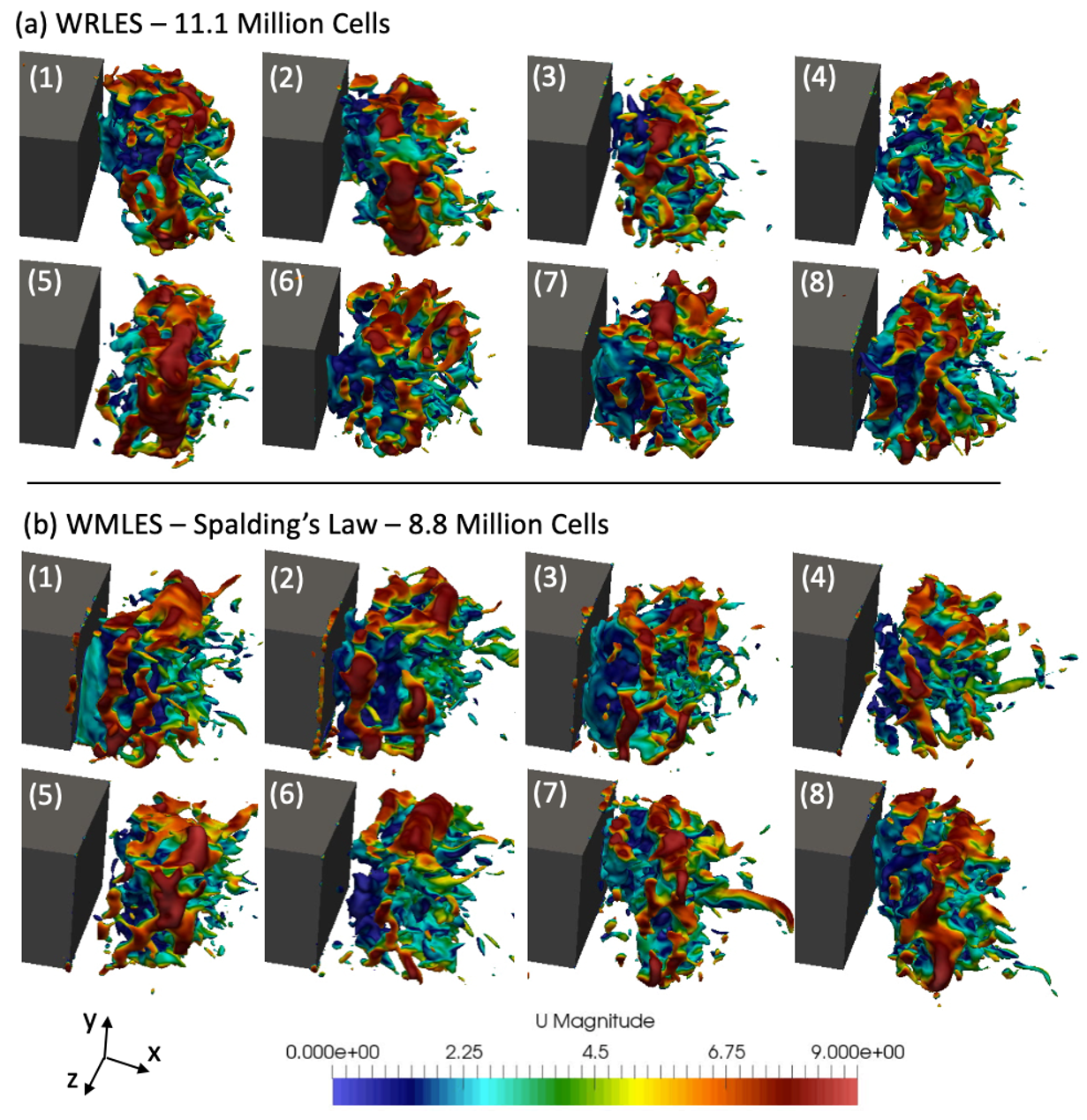}
\caption{Wake bimodality visualized using $C_P=-0.3$ iso-contour plots coloured by velocity magnitude for (a) WRLES and (b) 8.8 million cell WMLES using Spalding's law}
\label{fig:CpNeg03IsoContour}
\end{figure}

A three-dimensional visualisation of wake bimodality is possible by looking at iso-contour plots of $C_P$=-0.3 coloured by velocity magnitude. These plots are presented for WRLES in figure \ref{fig:CpNeg03IsoContour}(a) and for WMLES with Spalding's law in figure \ref{fig:CpNeg03IsoContour}(b) focusing on the Ahmed body's rear. The wake shifts in opposing spanwise directions for the two cases. In line with observations found in the computational studies of Lucas et al. \cite{RefWorks:218}, Rao et al. \cite{RefWorks:300}, Rao et al. \cite{RefWorks:297} and Dalla Longa et al. \cite{RefWorks:306}, the presented simulations resolve an instantaneously asymmetric wake that maintains its time-averaged toroidal structure and is simply tilted to a preferred side. This opposes the experimental observations made by Evrard et al. \cite{RefWorks:217}, Perry et al. \cite{RefWorks:216} and Pavia et al. \cite{RefWorks:329}, who, based on two-dimensional PIV planes, proposed that the instantaneously asymmetric wake is composed of single or multiple horseshoe vortices that, in the long time-average, coalesce into the closed toroidal structure characteristic of the time-averaged wake. In addition, the computational wake bimodality study of Dalla Longa et al. \cite{RefWorks:306} showed that switching from one bimodal state to the other is accompanied by the shedding of a large hairpin vortex that pulls the wake to the opposing off-axis position. This is not evident here. Instead, during a bimodal switch, the toroidal pressure structure merely \enquote{disconnects} from the rear base side to which it is initially tilted and then, following the bimodal event, reconnects to the rear base's opposing side. 

\begin{figure}
\centering
\graphicspath{ {Results/} }
\includegraphics[scale=0.35]{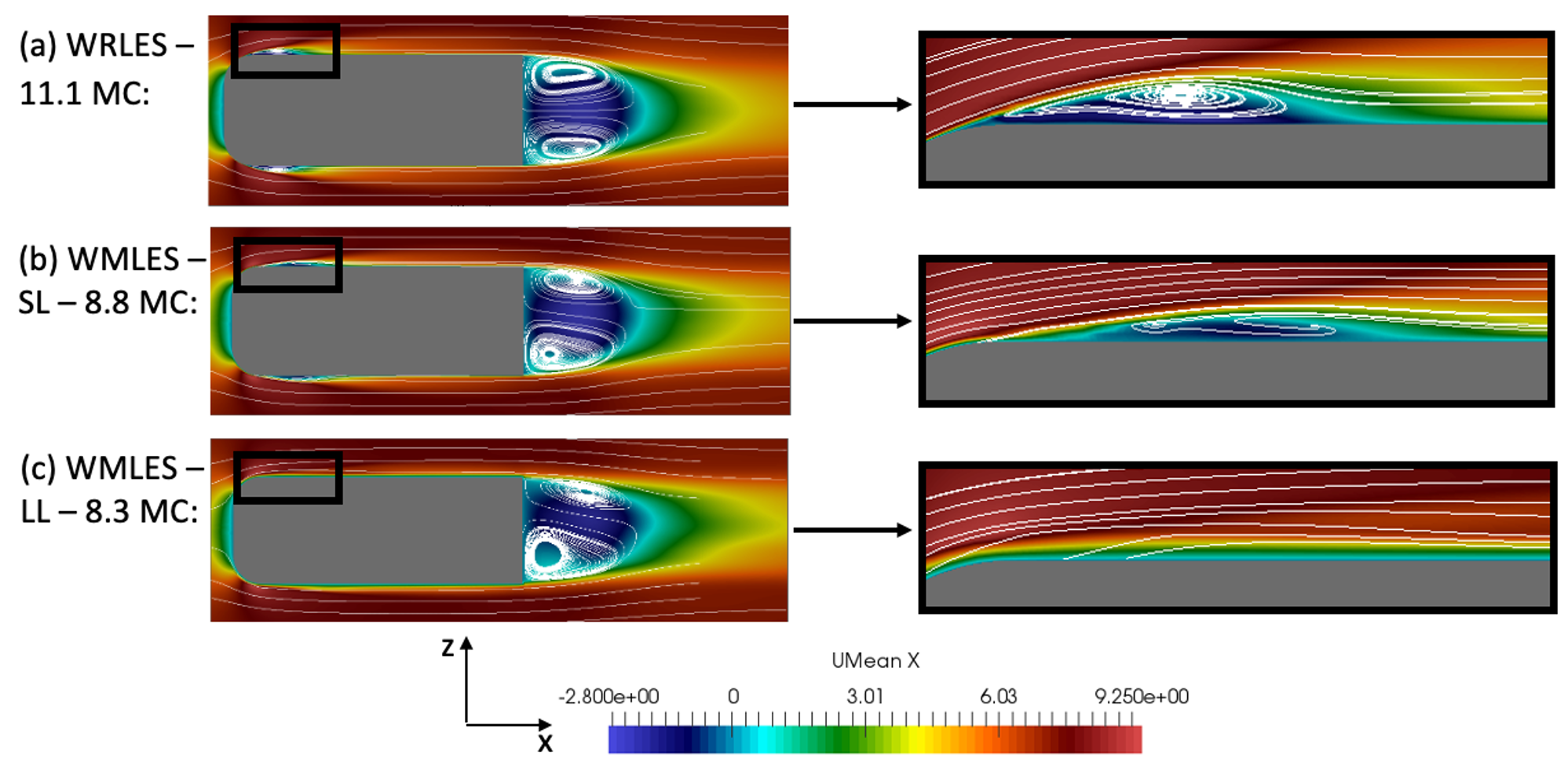}
\caption{Contour plots of the mean velocity's $x$-component with streamlines projected onto the $xz$ - middle plane passing through the Ahmed body for (a) WRLES, (b) 8.8 million cell (MC) WMLES with Spalding's law (SL) and (c) 8.3 million cell WMLES using the log-law (LL)}
\label{fig:UMeanx_ContourPlots}
\end{figure}

Overall, the occurrence of switches in the wake's horizontal asymmetry at non-regular time intervals (figure \ref{fig:BaseCPGradWRvsWMLES}(b)) illustrates bimodality's random nature and that it must be driven by stochastic processes introduced into the flow field as a result of the fluid's turbulent state. This corroborates Rigas et al. \cite{RefWorks:298}'s theoretical and experimental study of a three-dimensional wake formed by an axisymmetric bluff body. The investigation showed that the turbulent wake's dynamics, consisting of an infinite quantity of metastable states in the azimuthal direction and around a mean radial position that is determined by a supercritical pitchfork bifurcation originating in the laminar regime, can be accurately modelled by a two-dimensional nonlinear Langevin equation: the broken symmetry is accounted for by the deterministic component while the turbulent fluctuations forcing the wake to undergo a random walk are captured by the stochastic term \cite{RefWorks:298}. Furthermore, Dalla Longa et al. \cite{RefWorks:306} have already shown that WRLES can capture bimodality. They present one bimodal event for the Ahmed body in a $t^{*}$-window of $\sim$1400. Here, it is shown for the first time that WMLES with sufficient resolution can capture wake bimodality as well. Since the coarser WMLES set-up does not resolve bimodality with either of two equilibrium wall-models tested, it can be concluded that only when the near-wall region of the turbulent boundary layer, which consists of the viscous sub-layer and buffer layer, is modelled can bimodality still be resolved. The coarser 8.3 million cell setups namely treat the turbulent boundary layer's near-wall region as well as a portion of its outer region, which is made up of the log-layer, in a Reynolds-averaged sense, while the finer 8.8 million cell wall-model setups only model the turbulent boundary layer's near-wall dynamics. This suggests that the larger scale flow structures found in the outer region of the turbulent boundary layer, such as the vortices shed from the front separation bubble, contribute significantly to the flow intermittency/stochasticity that ultimately triggers a switch in the wake's horizontal orientation. Subsection \ref{SourceWakeBimodal} explores this idea further.

\subsection{Source of wake bimodality} \label{SourceWakeBimodal}

\begin{figure}
\centering
\graphicspath{ {Results/} }
\includegraphics[scale=0.18]{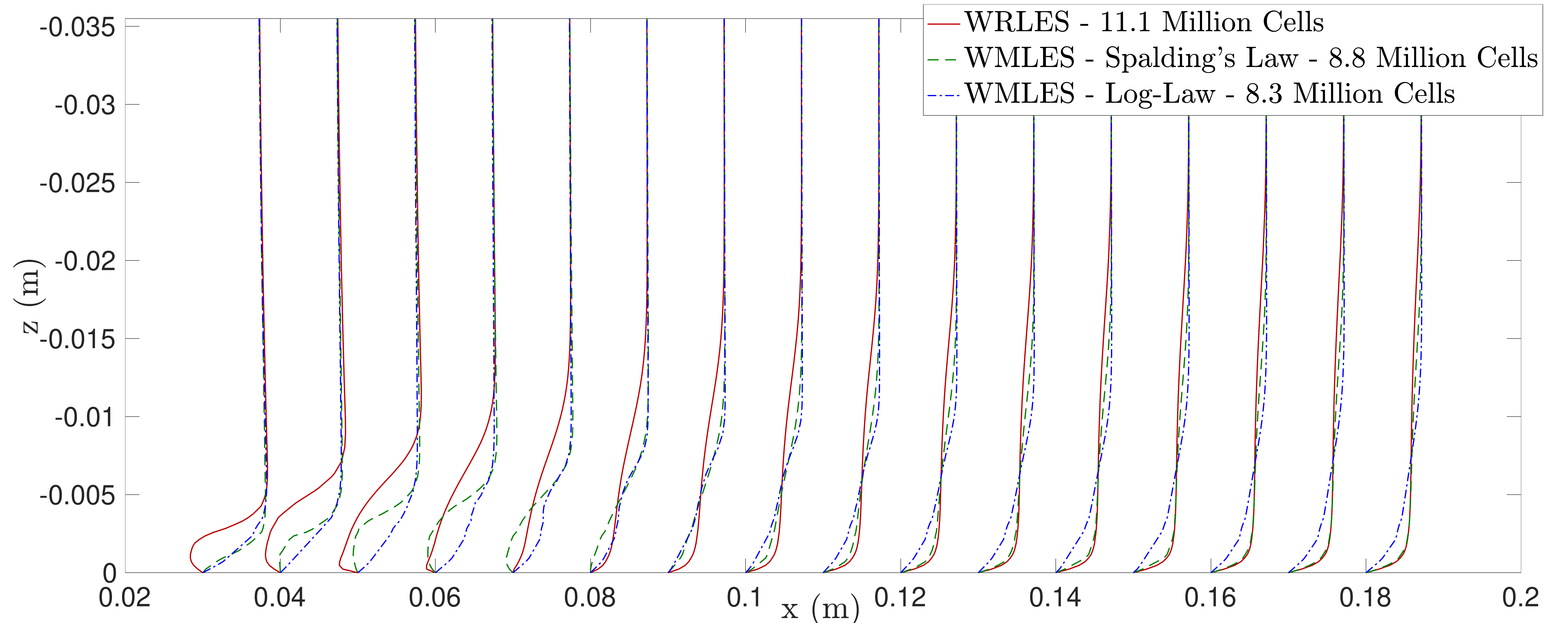}
\caption{Profiles of the mean velocity's $x$-component, $\bar{U}_x$, from the $xz$ - middle plane passing through the Ahmed body}
\label{fig:UMeanx_Profiles}
\end{figure}

\begin{figure}
\centering
\graphicspath{ {Results/} }
\includegraphics[scale=0.25]{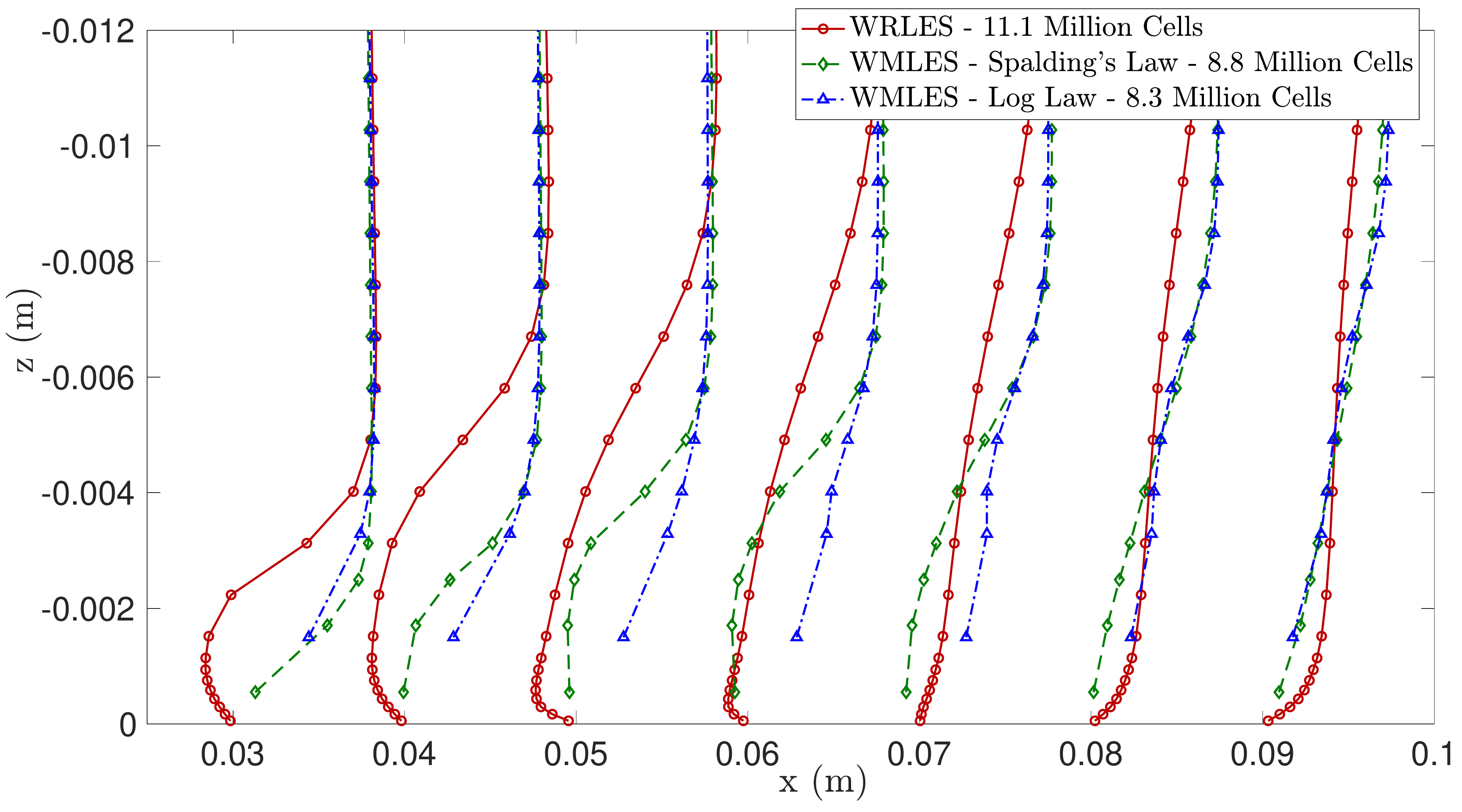}
\caption{Close-up of $x$-component mean velocity profiles in the vicinity of the front separation bubble. Markers denote cell centers }
\label{fig:CloseUP_UMeanx_Profiles}
\end{figure}

First, the mean flow structures captured in the three principal test cases are looked at. Attention is focused not on the wake, but on the front separation bubbles. These can be clearly seen using contour plots of the mean velocity's $x$-component with streamlines (figure \ref{fig:UMeanx_ContourPlots}). Visualization is performed on the same $xz$ - middle plane as the one used for figures \ref{fig:P_Contour_Switch} and \ref{fig:Umag_Contour_Switch}. Both the baseline 11.1 million cell WRLES (figure \ref{fig:UMeanx_ContourPlots}(a)) case and the 8.8 million cell WMLES simulation (figure \ref{fig:UMeanx_ContourPlots}(b)) capture the front separation bubble. As stated in the previous section, with WRLES the bubble is particularly distinct, while in the under-resolved wall-modelled near-wall region of the 8.8 million cell WMLES simulation, the bubble is slimmer and forms slightly downstream of the nose-to-side junction. These separation bubbles are present on the side surfaces and top surface of the Ahmed body. At the bottom side of the Ahmed body, ground proximity prevents detachment at the nose. Meanwhile, for the 8.3 million cell WMLES set-up (figure \ref{fig:UMeanx_ContourPlots}(c)), the extent to which the near-wall region is treated in a Reynolds-averaged sense is too large and no front separation bubble is detected on all four sides of the Ahmed body. By overlaying profiles of the mean velocity's $x$-component, $\bar{U}_x$, from the $xz$ - middle plane in the region of the front separation bubble (figure \ref{fig:UMeanx_Profiles}), it is possible to more precisely illustrate the difference in the mean flow structure captured among the three main test cases. A close-up of figure \ref{fig:UMeanx_Profiles} with cell center markers is also presented in order to more clearly depict the grid resolution's impact on the captured flow (separation) behavior (figure \ref{fig:CloseUP_UMeanx_Profiles}). Just downstream of the nose-to-side junction, where $x$ varies from 0.03m to 0.08m, the front separation is identifiable for both the baseline WRLES set-up and the 8.8 million cell WMLES case illustrated here with results for Spalding's law. In this area, $\bar{U}_x$ takes on non-positive values indicating recirculating flow. There are clear differences in the captured separation bubble though. Specifically, the flow separates from the Ahmed body side and then reattaches sooner for WRLES. Also, the separation bubble extends further into the wall-adjacent far-field for WRLES. The $\bar{U}_x$ profiles for WRLES and the 8.8 million cell WMLES simulation are coincident from $x$=0.12m onward, further validating the use of a mesh that under-resolves the near-wall flow behavior for Ahmed body aerodynamic simulations. Meanwhile, the $\bar{U}_x$ profiles for the 8.3 million cell WMLES set-up, shown here for the log-law, indicate attached flow throughout the domain downstream of the Ahmed body nose, inline with the observation that no front separation bubble is captured for this coarsened mesh construction. There are also clear discrepancies between the wall region's $\bar{U}_x$ profiles resolved with the 8.3 million cell mesh and the other two test cases. This illustrates that, although the coarse 8.3 million cell mesh is able to resolve horizontal wake asymmetry, it does not accurately capture the upstream turbulent boundary layers' profile and, hence, dynamics.

\begin{figure}
\centering
\graphicspath{ {Results/} }
\includegraphics[scale=0.37]{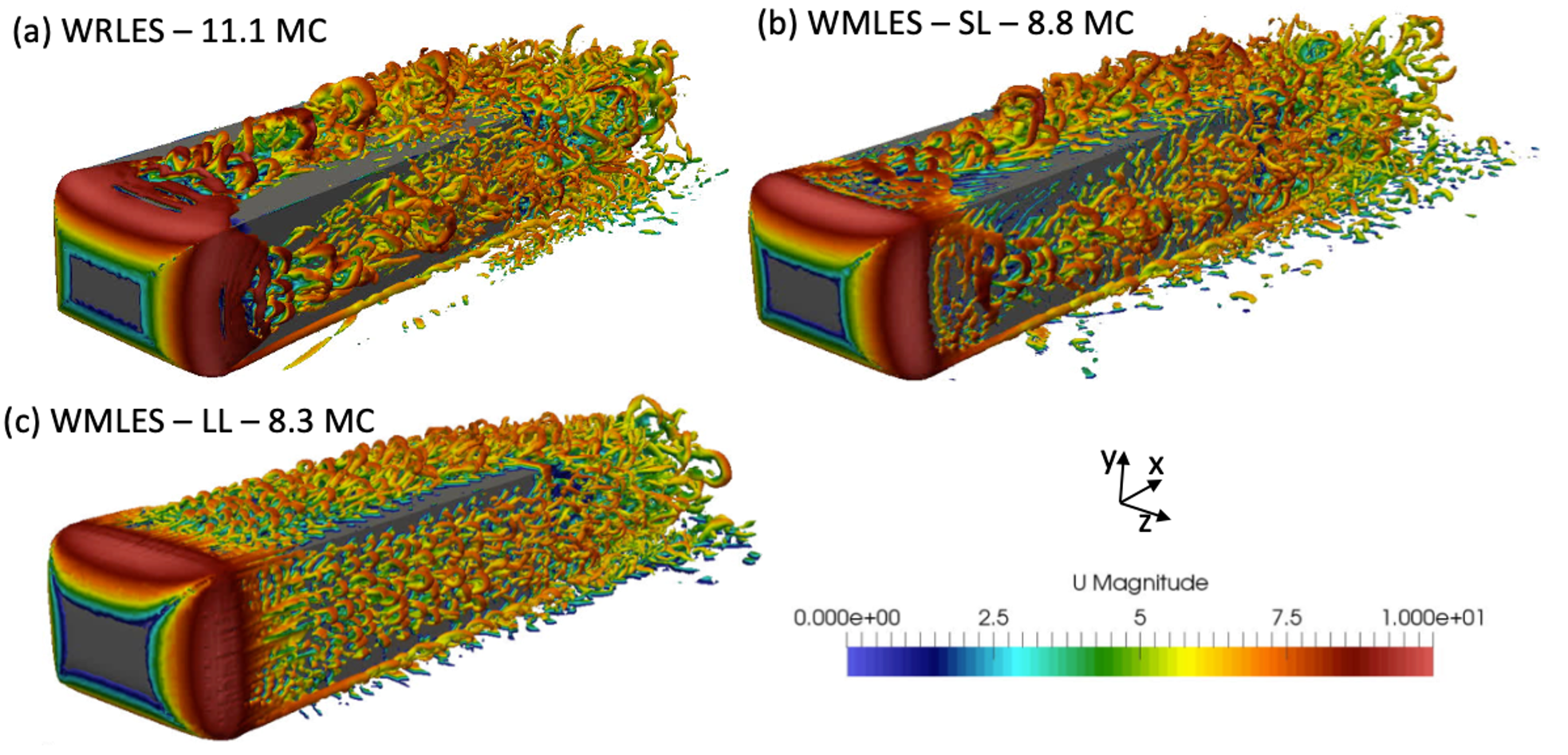}
\caption{Visualization of flow structures captured around the Ahmed body using iso-contours of $Q^{*}$=5.30 for (a) WRLES, (b) 8.8 million cell (MC) WMLES with Spalding's law (SL) and (c) 8.3 million cell WMLES using the log-law (LL)}
\label{fig:QStar530IsoContour}
\end{figure}

\begin{figure}
\centering
\graphicspath{ {Appendix/} }
\includegraphics[scale=0.37]{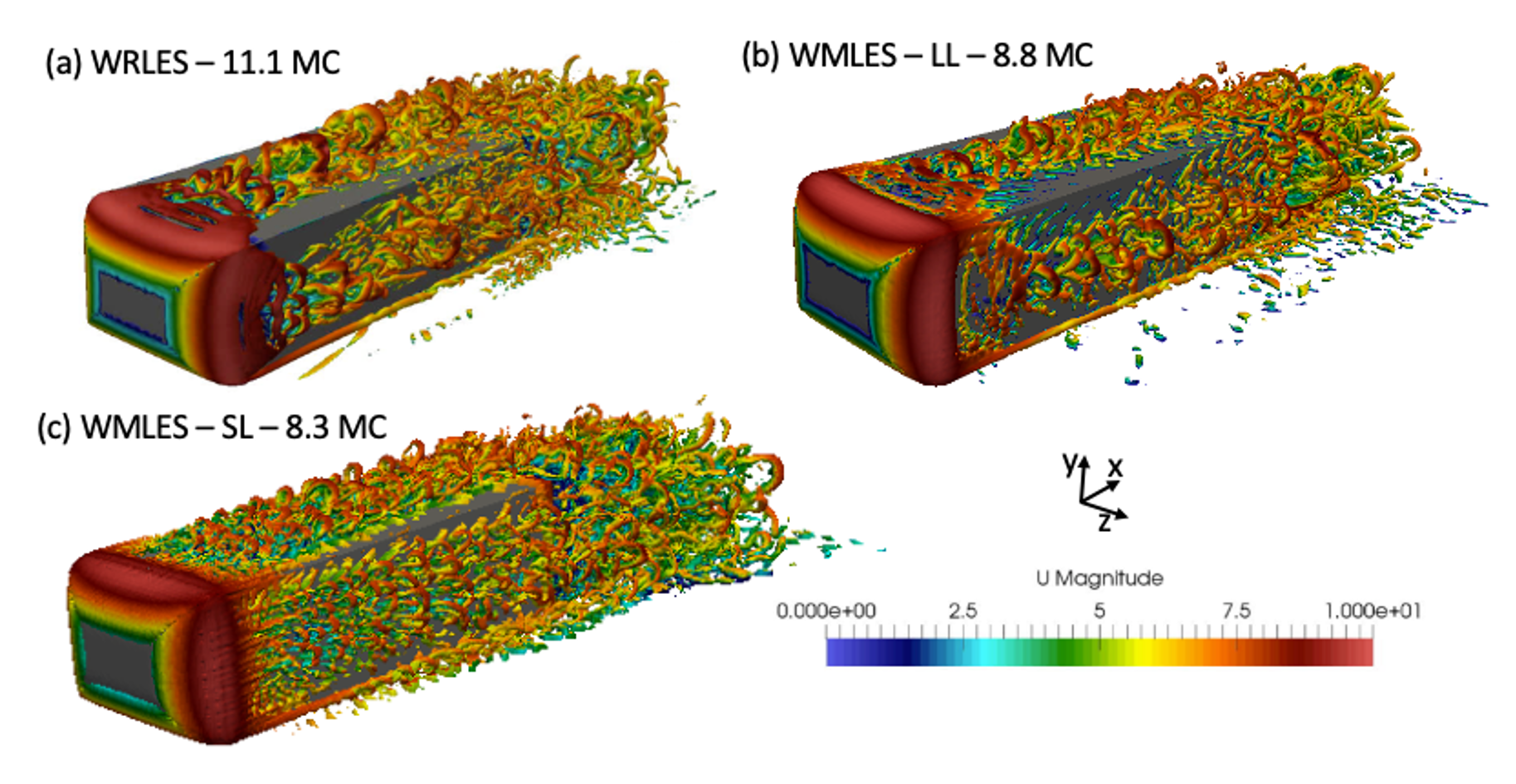}
\caption{Visualization of flow structures captured around the Ahmed body using iso-contours of $Q^{*}$=5.30 for (a) WRLES, (b) 8.8 million cell (MC) WMLES with the log-law (LL) and (c) 8.3 million cell WMLES using Spalding's law (SL)}
\label{fig:QStar530IsoContour_Appendix}
\end{figure}

Next, the instantaneous flow structures resolved around the Ahmed body in the three principal simulations are analysed. Three-dimensional visualization of these features is accomplished by looking at iso-contour plots of $Q^{*}$ coloured by velocity magnitude (figure \ref{fig:QStar530IsoContour}). The selected value is $Q^{*}=5.30$, thereby enabling the identification of structures with high rotation and low pressure cores. Focus is again placed on the area upstream of the wake, particularly the nose-to-side junction and the region just downstream of this junction. In figure \ref{fig:QStar530IsoContour}(a) for the 11.1 million cell WRLES simulation and figure \ref{fig:QStar530IsoContour}(b) for the 8.8 million cell WMLES test case, shown here with Spalding's law applied, similar vortex structures are shed from the Ahmed body's nose. These shed structures originate from the previously identified prominent front separation bubbles (figure \ref{fig:UMeanx_ContourPlots}(a) and \ref{fig:UMeanx_ContourPlots}(b)) and develop into large hairpin vortices that are convected along the Ahmed body's sides to the rear. Such flow features are not present in figure \ref{fig:QStar530IsoContour}(c) for the 8.3 million cell WMLES set-up illustrated here with results for the log-law. The test case's inability to capture front separation bubbles (figure \ref{fig:UMeanx_ContourPlots}(c)) results in its deficiency to resolve the large hairpin vortices that are generated by the bubbles. Instead, a streaky structure of smaller scale hairpin vortices manifest themselves along the Ahmed body's sides. The Appendix presents a corresponding plot for figure \ref{fig:QStar530IsoContour} where Spalding's law and the log-law of the wall are applied on reversed mesh order (figure \ref{fig:QStar530IsoContour_Appendix}). It further illustrates bimodality's sensitivity to the upstream near-wall Ahmed body mesh resolution.  

\begin{figure}
\centering
\graphicspath{ {Results/} }
\includegraphics[scale=0.35]{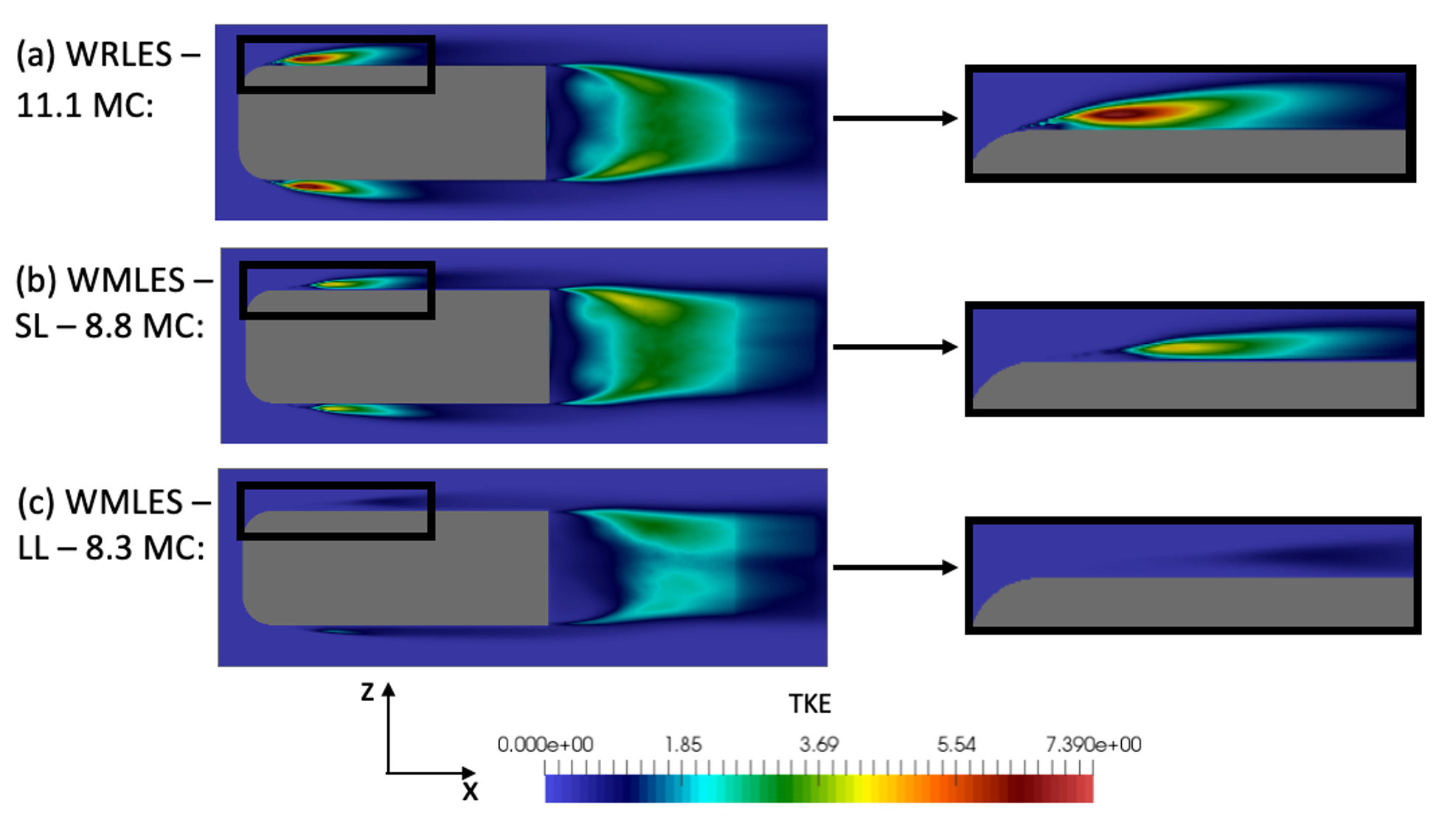}
\caption{Contour plots of TKE projected onto the $xz$ - middle plane passing through the Ahmed body for (a) WRLES, (b) 8.8 million cell (MC) WMLES with Spalding's law (SL) and (c) 8.3 million cell WMLES using the log-law (LL)}
\label{fig:TKE_ContourPlots}
\end{figure}

\begin{figure}
\centering
\graphicspath{ {Results/} }
\includegraphics[scale=0.18]{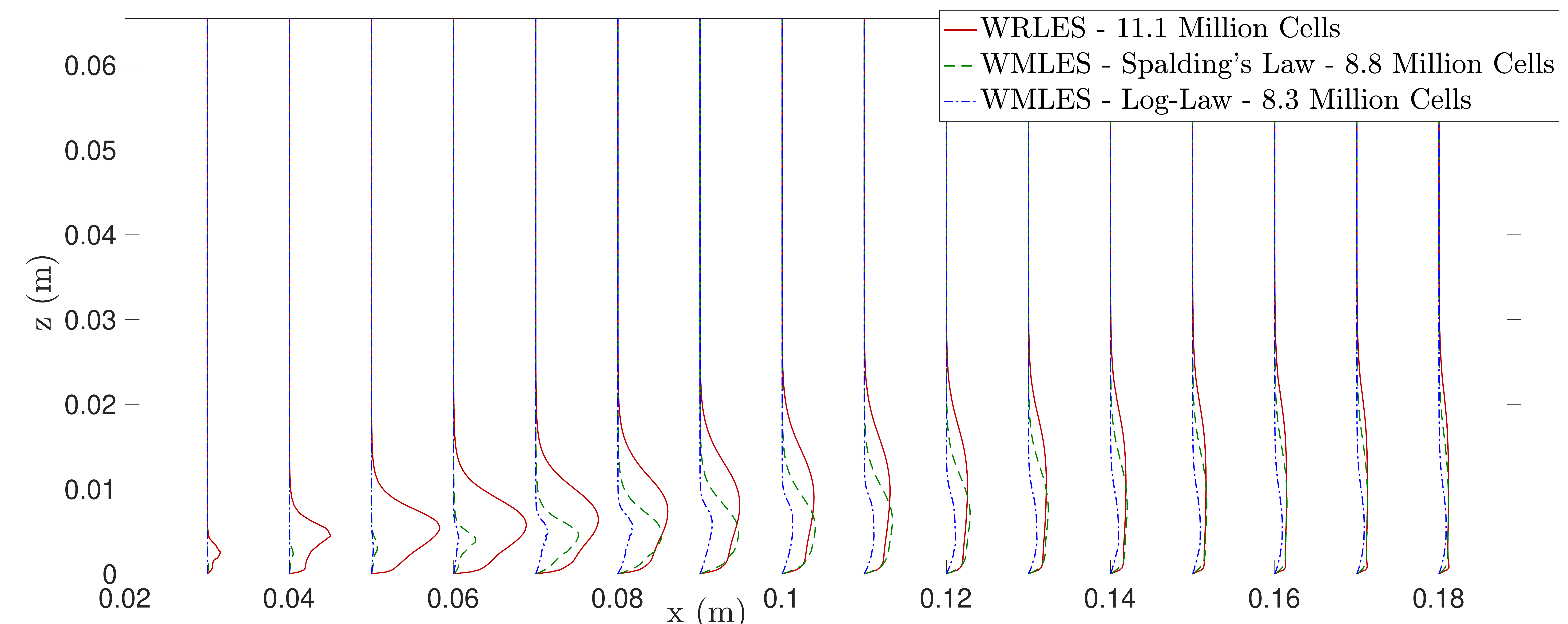}
\caption{Profiles of the TKE from the $xz$ - middle plane passing through the Ahmed body}
\label{fig:TKE_Profiles}
\end{figure}

The distinct difference in instantaneous flow structures simulated directly translates to a difference in the level of flow intermittency generated at the Ahmed body nose. By looking at the turbulent kinetic energy (TKE) that is produced from the flow structures around the Ahmed body, the critical role that the front separation bubbles play in generating flow intermittency through the shedding of large hairpin vortices can be clearly discerned. Figure \ref{fig:TKE_ContourPlots} presents contour plots of this TKE on the $xz$ - middle plane passing through the Ahmed body. Significant amounts of TKE are produced at the front separation bubble for both the 11.1 million cell WRLES set-up (figure \ref{fig:TKE_ContourPlots}(a)) and the 8.8 million cell WMLES simulation illustrated here with data for Spalding's law (figure \ref{fig:TKE_ContourPlots}(b)). The TKE level is particularly high for WRLES. It is, however, negligible for the highly under-resolved 8.3 million cell WMLES set-up depicted here with log-law results (figure \ref{fig:TKE_ContourPlots}(c)). This is simply because the coarse 8.3 million cell mesh fails to capture the flow structure responsible for this TKE generation, the front separation bubble. Profile cuts of the TKE (figure \ref{fig:TKE_Profiles}), just as was done for $\bar{U}_x$, further illustrate the stark difference in TKE levels captured by the two meshes that are able to resolve wake bimodality and the grid that is not capable of simulating wake bimodality. Non-negligible amounts of TKE spread out into the wall-adjacent far-field furthest for WRLES. The TKE level also builds-up earliest for WRLES with the TKE profiles for WRLES matching those of the 8.8 million cell WMLES simulation from about $x$=0.15m onward.

\begin{figure}
\centering
\graphicspath{ {Results/} }
\includegraphics[scale=0.31]{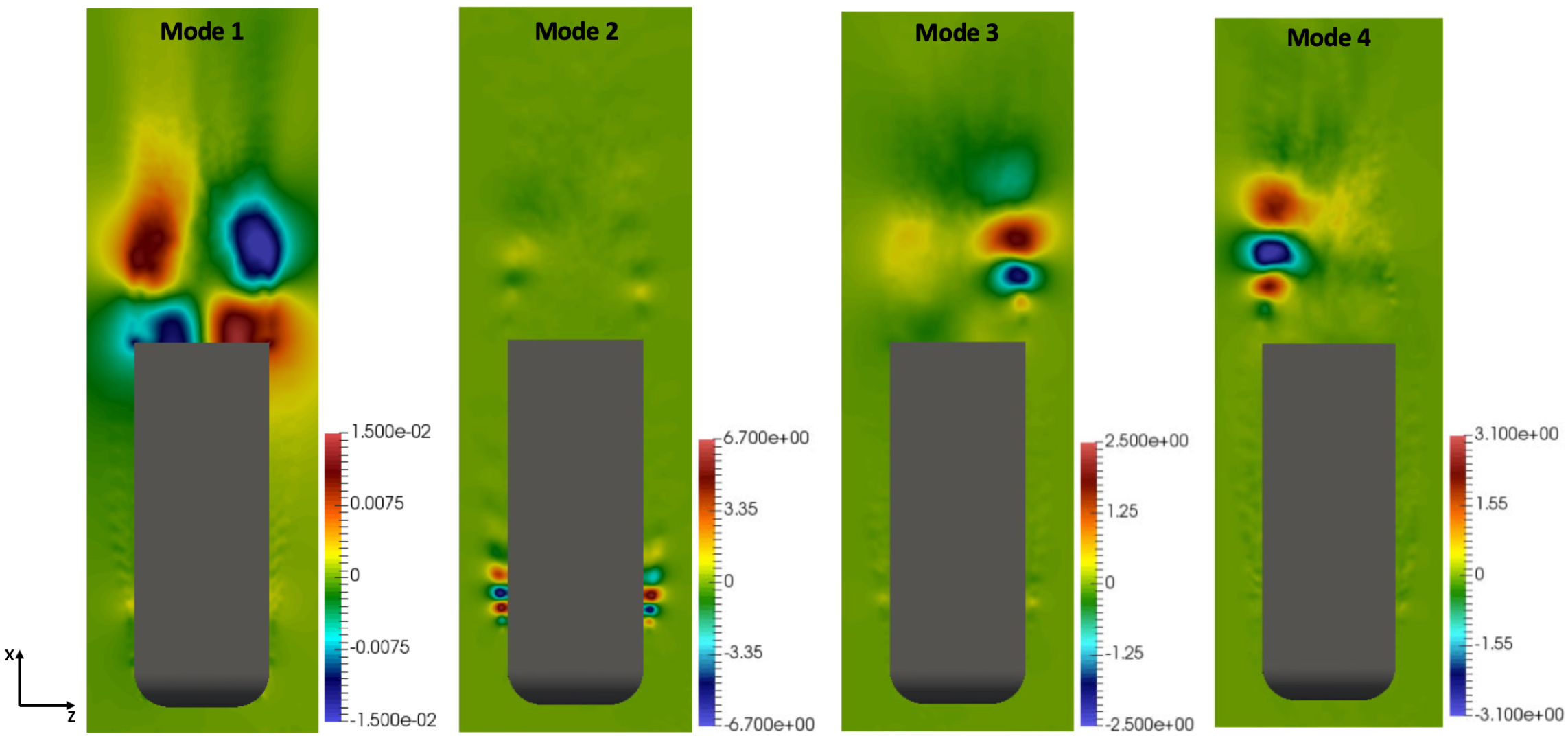}
\caption{POD modes from pressure data on the $xz$ - middle plane passing through the Ahmed body for the 8.8 million cell WMLES set-up using Spalding's law}
\label{fig:POD_Modes}
\end{figure}

Having identified that front separation bubbles shed large hairpin vortices and, thereby, generate significant amounts of TKE, proper orthogonal decomposition (POD) is now applied on pressure data from the $xz$ - middle plane to illustrate the coherent structures captured in each simulation. This is done to further exemplify the importance of the front separation bubble in providing the flow intermittency necessary to trigger a bimodal event. Figure \ref{fig:POD_Modes} depicts the POD modes for the 8.8 million cell WMLES simulation using results for Spalding's law. Mode 1 represents the bimodal mode. It is a static mode and, therefore, does not come with a 90$^{\circ}$ out-of-phase mode that, when coupled to it, produces a convective motion. The remaining three modes do come in pairs though, illustrating that they are fundamentally convective in nature. Mode 2 represents the front separation bubble and its shedding dynamics. Modes 3 and 4 are the asymmetric rear vortex shedding modes, each associated with a different horizontal orientation of the wake. POD modes similar in structure are computed from the pressure field snapshots of the WRLES case and may be found in the Appendix (figure \ref{fig:POD_Modes_Appendix}). This is no surprise. The set-up captures the front separation bubble (albeit slightly further upstream in position) as well as its associated shedding dynamics, and, thus, resolves wake bimodality with which two asymmetric rear vortex shedding modes are coupled. Applying POD to the pressure field snapshots obtained from the 8.3 million cell WMLES set-up produces a far more restricted array of modes. Specifically, only one clearly identifiable coherent structure is detected. It is a single asymmetric rear vortex shedding mode whose orientation is determined by the spanwise side to which the wake is tilted during the simulation time-span. No mode representative of the front separation bubble's shedding dynamics is computed and, in turn, no bimodal static mode is present. This reiterates the importance of the front separation bubble's dynamics in producing the flow intermittency needed to trigger a bimodal event. 

\section{Conclusion} \label{Conclusion}
A $Re_H=33,333$ flow past a 1/4 scale squareback Ahmed body is simulated using LES. The objective is to study the horizontally bimodal wake aft of the Ahmed body. A particular focus is placed on the flow structures likely to contribute significantly to triggering a bimodal wake switching event. WRLES and WMLES simulations are carried out. The two equilibrium wall-models explored are Spalding's law of the wall and the log-law of the wall. By using unstructured meshes, a WRLES mesh, capable of capturing wake bimodality, that is 80\% smaller in cell count than previous fully block-structured meshes used in computational wake bimodality studies is created. For the first time, it is shown that WMLES is able to resolve wake bimodality as well. Specifically, by applying both wall-models on a WMLES mesh that only treats the inner region of the turbulent boundary layer in a Reynolds-averaged sense, it is possible to still simulate wake bimodality. Not only does this reduce the numerical resources needed to capture the bimodal behavior of the rear wake by a further 21\%, but it illustrates that the larger scale flow structures found in the outer region of the turbulent boundary layer must contribute appreciably to the flow intermittency that, following convection to the Ahmed body's rear, forces the wake into opposing off-axis locations. By looking at the larger scale flow structures present in the set-ups capable of simulating wake bimodality and comparing these to the structures resolved in the wake bimodality deficient case, it is evident that the front separation bubbles present at the Ahmed body's nose generate particularly high levels of TKE by shedding large hairpin vortices. Without the resolution of these front separation bubbles and their associated flow intermittency, wake bimodality is not captured. This suggests that effective control strategies to suppress wake bimodality and re-centre the wake, thereby improving bluff-body drag, should act on the front separation bubbles. These results also may indicate that streamlined squareback bluff-bodies, such as the Renault Kangoo in the study by Bonnavion et al. \cite{RefWorks:314} or the Citro\"{e}n Berlingo in the study by Bonnavion et al. \cite{RefWorks:317}, might first present wake bimodality once the vehicle is yawed and/or pitched because only in these conditions do prominent separation bubbles manifest themselves at the vehicle's front. 

\section*{Acknowledgemeants}
The authors would like to thank Siemens PLM for providing a Star-CCM+ license and user-support where needed. Financial support from the Engineering and Physical Sciences Research Council (EPSRC) and Nissan is also greatly appreciated. This work benefited immensely from computational power provided by the Imperial College of London HPC cluster, CX2, and the UK's leading computing facility, ARCHER, as well.

\section*{Appendix}

\begin{figure}
\centering
\graphicspath{ {Appendix/} }
\includegraphics[scale=0.44]{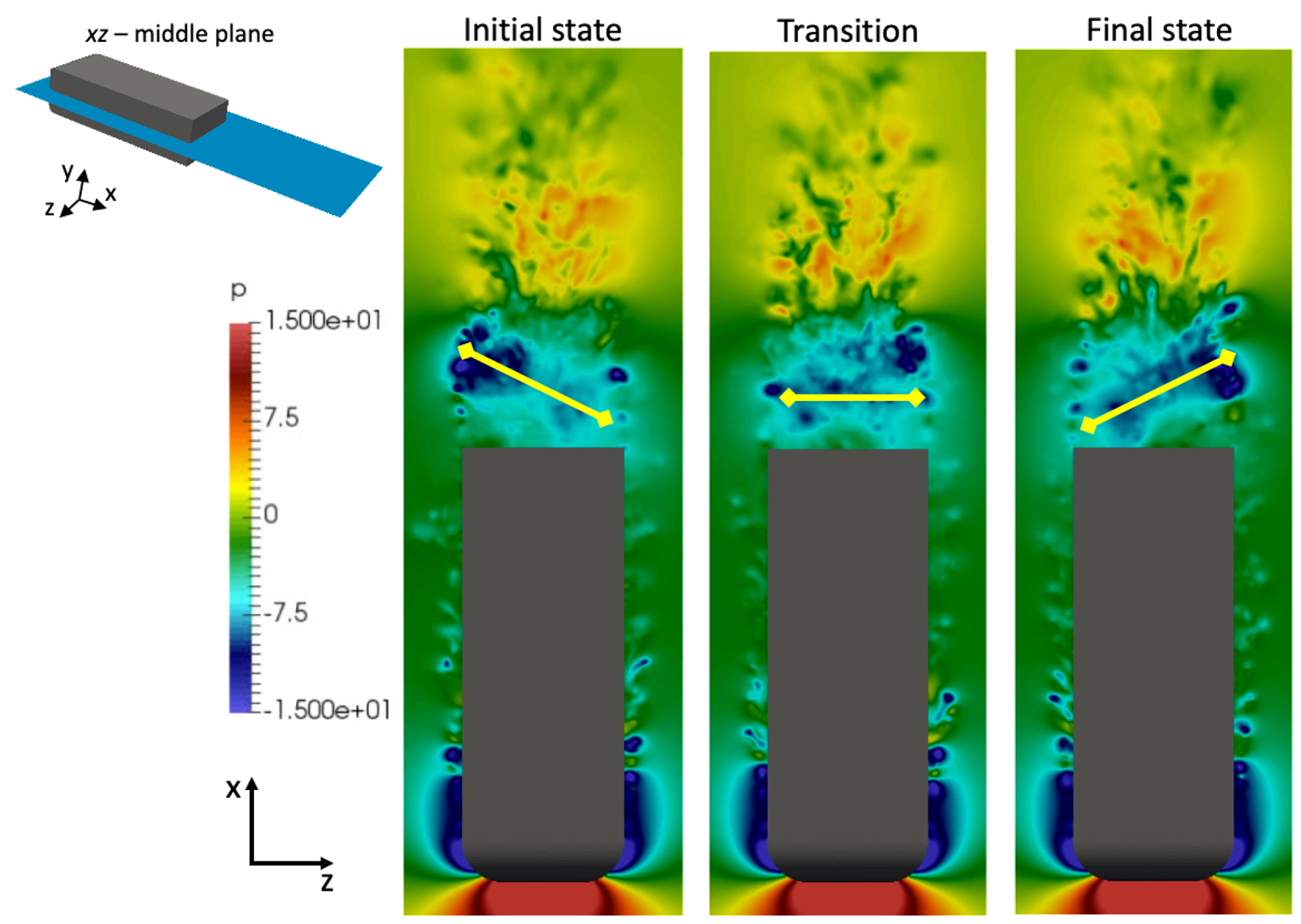}
\caption{Wake bimodality visualized using the instantaneous pressure field projected onto the $xz$ - middle plane passing through the Ahmed body for the 8.8 million cell WMLES case that uses Spalding's law}
\label{fig:P_Contour_Switch_WMLES_Appendix}
\end{figure}

\begin{figure}
\centering
\graphicspath{ {Appendix/} }
\includegraphics[scale=0.31]{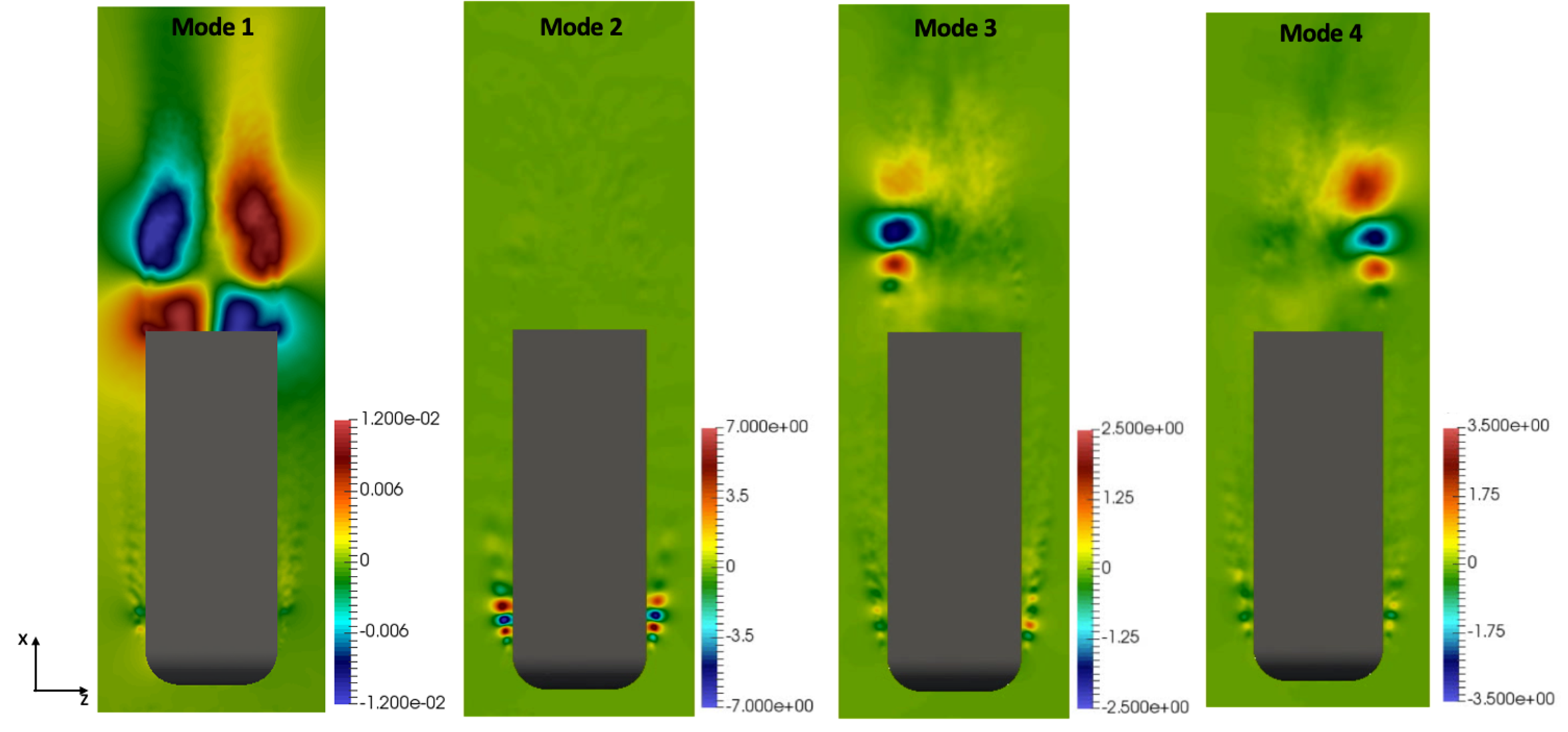}
\caption{POD modes from pressure data on the $xz$ - middle plane passing through the Ahmed body for the 11.1 million cell WRLES set-up}
\label{fig:POD_Modes_Appendix}
\end{figure}

\clearpage

\bibliography{References}

\end{document}